\documentclass[aps,prb,twocolumn,reprint,floatfix,nofootinbib]{revtex4-2}
\usepackage{graphicx}
\usepackage{amsmath,amssymb,mathrsfs,amsthm}
\usepackage[pdftex, colorlinks=true, linkcolor=blue, citecolor=blue, urlcolor=blue]{hyperref}
\usepackage{bbold}
\usepackage{times}
\usepackage{mathtools}

\usepackage{lipsum}
\usepackage{svg}
\usepackage{comment}
\usepackage[normalem]{ulem}
\usepackage{siunitx}
\usepackage{color}

\begin{document}

\title{Disorder-enhanced transport in a chain of lossy dipoles strongly coupled to cavity photons}

\author{Thomas F.\ Allard}
\affiliation{Universit\'e de Strasbourg, CNRS, Institut de Physique et Chimie des Mat\'eriaux de Strasbourg, UMR 7504, F-67000 Strasbourg, France}
\author{Guillaume Weick}
\affiliation{Universit\'e de Strasbourg, CNRS, Institut de Physique et Chimie des Mat\'eriaux de Strasbourg, UMR 7504, F-67000 Strasbourg, France}

\begin{abstract}

We study the interplay between disorder and light-matter coupling by considering a disordered one-dimensional chain of lossy dipoles coupled to a multimode optical cavity, through a microscopically derived Hamiltonian. 
Such a system, hosting polaritonic excitations, may be realized experimentally in a wide range of platforms under strong light-matter coupling. 
By analyzing both the eigenspectrum and the driven-dissipative transport properties of our system, we find that in the strong-coupling regime, increasing disorder leads almost uncoupled dark states to acquire a photonic part, allowing them to inherit polaritonic long-range transport characteristics.
Crucially, we show that this disorder-enhanced transport mechanism is increasingly noticeable when the considered dipoles are lossier.

\end{abstract}

\maketitle

\section{Introduction}

Over the past several years, strong coupling of matter excitations with confined electromagnetic modes has been shown to significantly modify material properties \cite{EbbesenReview2021}.
In particular, it has been proposed both
theoretically \cite{Feist2015,Schachenmayer2015,Gonzalez-Ballestero2016,Hagenmuller2017,Botzung2020,Chavez21,Dubail22,Gera_arXiv21,Engelhardt2022,Patton_arXiv22,Yang2022,Ribeiro2022} and experimentally \cite{Lerario2014,Orgiu2015,Zhong2017,Lerario2017,Rozenman2018,Hou2020,Nagarajan2020,Pandya2022,Xu_arXiv22,Schwartz_arXiv22} to use the strong light-matter coupling between emitters and cavity photons to modify transport properties and induce long-range transport, especially in disordered, molecular systems.
This cavity-enhanced transport effect can be understood from the fact that in the strong-coupling regime, the matter excitations hybridize with cavity photons, leading to collective polaritonic excitations that are delocalized throughout the whole system.

Notably, the interplay between strongly coupled light-matter systems and disorder has been shown to lead to unexpected phenomena, such as a cavity-protection effect, namely that polaritonic states are more robust than others against disorder \cite{Houdre1996,Michetti_2005}.
Such a behavior goes beyond the well-known theory of Anderson localization with short-range interaction, which states that in one-dimensional (1d) systems, disorder makes all the eigenstates exponentially localized, thereby suppressing transport \cite{Anderson1958,Mirlin_RevModPhys}.
For long-range interaction, however, the situation becomes highly nontrivial, since the 1d system can host extended states \cite{Levitov1990,Mirlin1996,Malyshev2003,Malyshev2005,Deng2018}.
Since polaritons are partly mediated by cavity photons, hence through an effective long-range coupling, they are less affected by the suppression of transport arising from Anderson localization.
Interestingly, dark states, i.e., matter states that are only weakly hybridized to photonic degrees of freedom, can also show nontrivial transport properties induced by the strong-coupling regime \cite{Gonzalez-Ballestero2016,Botzung2020,Dubail22,Engelhardt2022,Pandya2022}.

Alongside the above-mentioned cavity-protection and cavity-enhanced transport effects, recent works \cite{Chavez21,Dubail22,Engelhardt2022} have shown that the photonic cavity can also lead to an improvement of the transport characteristics when increasing the disorder strength, instead of an expected suppression.
This counterintuitive phenomenon was theoretically unveiled by considering a disordered single-mode Tavis-Cummings model, namely a chain of emitters coupled to a single cavity mode through a spatially homogeneous coupling constant.
Such enhanced transport induced by disorder has been very recently observed in experiments \cite{Cohn_2022,Brantut_arXiv2022,Long_arXiv2022}.

\begin{figure}[tb]
 \includegraphics[width=\linewidth]{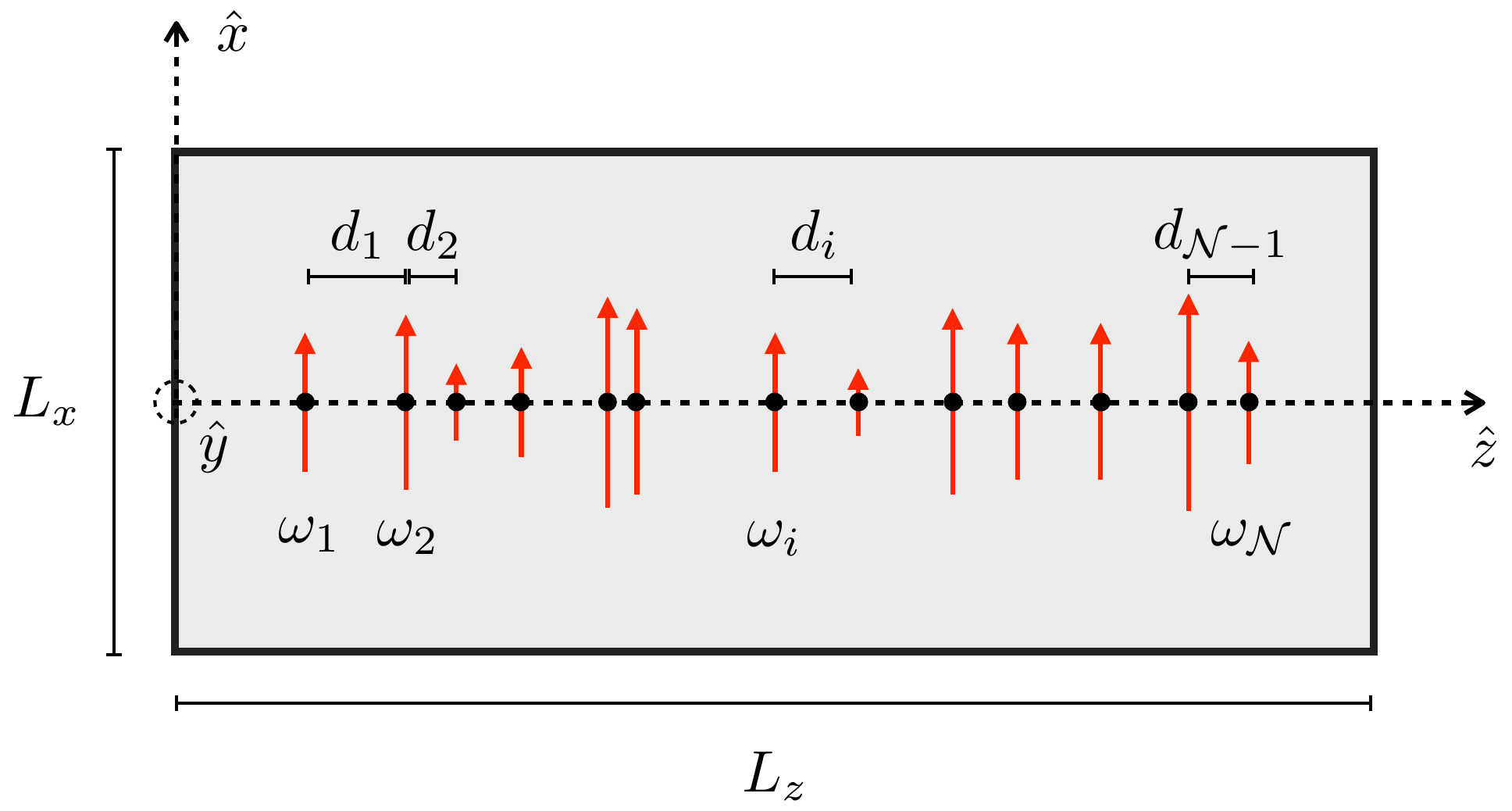}
 \caption{Sketch of a disordered chain of $\mathcal{N}$ oscillating dipoles polarized along the $x$ axis and arranged along the $z$ direction of a cuboidal cavity of linear sizes $L_x$, $L_y$, and $L_z$. Each dipole of frequency $\omega_i$, depicted by a red arrow, is located at a distance $d_i$ from its neighbor to the right.}
 \label{fig:sketch_cavity}
\end{figure}

The interplay between disorder and strong light-matter coupling is therefore highly nontrivial, and its understanding is of primary importance, since disorder is always present in experimental setups.
Here, we address such interplay by considering a finite disordered 1d chain of $\mathcal{N}$ oscillating dipoles placed inside a mirror cavity (see Fig.~\ref{fig:sketch_cavity}).
We keep our model general such that these dipoles represent a wide range of physical systems whose main coupling mechanism is of dipolar nature, and which are not dominated by quantum effects. These range from plasmonic, dielectric or SiC nanoparticles \cite{Krenn1999,Maier2003,Koenderink2007,Markel,Crozier07,Koenderink2007,Apuzzo2013,Slobozhanyuk2015,Wang2018b,Zhang2020} to magnonic microspheres \cite{Pirmoradian2018,Rameshti_arXiv2021} or macroscopic microwave helical antennas \cite{Mann2018,Mann2020}.
Our model can also describe molecular and semiconductor excitons \cite{Feist2015,Yuen-Zhou2016}, for which most of the recent experimental approaches on disordered polaritonic systems were performed \cite{Lerario2014,Orgiu2015,Zhong2017,Lerario2017,Rozenman2018,Hou2020,Nagarajan2020,Pandya2022,Xu_arXiv22,Schwartz_arXiv22}, as well as cold atoms \cite{Browaeys2016,Barredo2016,Perczel2017,Wang2018a} where controllable frequency disorder has been achieved \cite{Brantut_arXiv2022,Long_arXiv2022}, and which, in the single excitation manifold, behave as classical dipoles \cite{Asenjo-Garcia2019}.

We consider dipoles that are coupled to each other through an all-to-all short-range quasistatic dipole-dipole interaction, and, importantly, we go beyond the usually considered single mode Tavis-Cummings model by studying a microscopically derived minimal coupling Hamiltonian \cite{Downing2019,Downing2021} which couples the dipoles to a multimode cavity.
Since the majority of materials used in experimental polaritonics, such as Fabry-Pérot or plasmonic cavities, are intrinsically multimodal, it is of fundamental importance to go beyond single-mode models  \cite{Tichauer2021,Ribeiro2022}.
We take into account disorder both in the individual resonance frequencies of each dipole, modeling possible inhomogeneities and in the interdipole distances, accounting for positioning uncertainties. 
Crucially, unlike the recent studies of Refs.~\cite{Chavez21,Dubail22,Engelhardt2022}, we study transport through an out-of-equilibrium driven-dissipative scenario, which permits us to analyze the propagation of polaritonic excitations along the chain as a function of the driving frequency, and to study the effect of dissipation, taking into account losses both in the dipoles and in the mirror cavity. 

Our model allows us to show that disorder-enhanced transport can be readily explained as a result of an hybridization between dark and polaritonic states, which is induced by the bandwidth increase led by frequency disorder. 
This disorder-induced hybridization enables dark states to inherit polaritonic properties, namely to take advantage of cavity-enhanced transport.
Importantly, we show that several different transport regimes are found for chains of large-enough size, so that cavity- and disorder-enhanced transport only occur at intermediate distances. 
Indeed, the short-range propagation of the dark states, which results from the nearest-neighbor quasistatic dipole-dipole coupling only, cannot be enhanced through polaritonic hybridization, and only the long-range propagation can profit from cavity- and disorder-enhanced transport.
Moreover, no enhancement is found over very long distances.
Crucially, the consideration of dipolar losses is proved to be of great importance, since the latter emphasizes the aforementioned effects of cavity- and disorder-enhanced transport.

The paper is organized as follows: Section \ref{sec:Model} is dedicated to the presentation of our microscopic model Hamiltonian of a disordered chain of dipoles coupled to a cuboidal photonic cavity. 
In Sec.~\ref{sec:Spectrum}, we study the localization properties of the system from an eigenspectrum analysis, both in the case of ordered and disordered chains, where we discuss the mixing between dipolar and photonic degrees of freedom induced by the interplay between strong light-matter coupling and disorder.
In Sec.~\ref{sec:Transport}, we study the transport along the chain in a driven-dissipative scenario, considering both lossy dipoles and lossy cavity mirrors. 
We first study the effects of strong coupling in an ordered chain, and then the interplay between disorder and cavity-enhanced transport in a disordered chain.
Finally, we summarize our results and draw conclusions in Sec.~\ref{sec:Conclusion}. 

In  Appendix \ref{sec:Diagonalization} we reproduce for 
the sake of self-containedness results for the case of an ordered chain \cite{Downing2021}, while in Appendix \ref{sec:RWA} we discuss some approximations adopted in our model.
In Appendixes \ref{sec:multifractality}, \ref{sec:Stationary transport details}, \ref{sec:Cavity losses}, and \ref{sec:Absorption}, we complement the discussion of the main text by studying the precise nature of the semilocalized eigenstates when increasing the system size, giving technical details of our transport simulations, investigating the influence of cavity losses, and presenting absorption spectra obtained from transport calculations, respectively.

\section{Disordered chain of dipoles coupled to cavity photons}
\label{sec:Model}

To describe a disordered chain of $\mathcal{N}$ dipoles coupled to cavity photons, we rely on the model recently developed in Refs.~\cite{Downing2019, Downing2021} for an ordered chain, and supplement it by including disorder both in the interdipole spacings $d_i$ and in the individual resonance frequencies $\omega_i$ (see Fig.~\ref{fig:sketch_cavity}).
Our model is described by the Hamiltonian
\begin{equation}
    H = H_{\mathrm{dp}} + H_{\mathrm{ph}} + H_{\mathrm{dp}\textrm{-}\mathrm{ph}}.
\label{eq:Hamiltonian}
\end{equation}
The first term on the right-hand side (r.h.s.) of Eq.~\eqref{eq:Hamiltonian} accounts for the dipolar degrees of freedom and reads \cite{Downing2017_Retardation} 
\begin{equation}
    H_{\mathrm{dp}} = \sum_{i=1}^\mathcal{N} \hbar\omega_i b_{i}^{\dagger} b_{i}^{\phantom{\dagger}} + \sum_{i=1}^{\mathcal{N}-1} \sum_{j=i+1}^\mathcal{N} \hbar\Omega_{ij}\left( b_i^\dagger b_j^{\phantom{\dagger}} + b_{i}^{\phantom{\dagger}}b_{j}^{\dagger} \right),
\label{eq:H_dp}
\end{equation}
where counter-rotating terms in the quasistatic dipole-dipole coupling have been ignored.
A treatment of the Hamiltonian \eqref{eq:Hamiltonian} without using the rotating wave approximation is discussed in Appendix~\ref{sec:RWA}.
In the above dipolar Hamiltonian \eqref{eq:H_dp}, the bosonic operator $b_i^{\phantom{\dagger}}$ ($b_i^{\dagger}$) annihilates (creates) a dipolar excitation, polarized along the $x$ axis and with resonance frequency $\omega_i$ on site $i \in [1,\mathcal{N}]$.
The latter individual resonance frequencies are uncorrelated random variables distributed uniformly within an interval $[\omega_0 - W/2, \omega_0 + W/2]$, hence with an average resonance frequency being $\omega_0$.
The dipolar coupling strength $\Omega_{ij}$ between two dipoles on sites $i$ and $j$ separated by a distance $r_{ij}=\sum_{l=i}^{j-1}d_l$ is given by
\begin{equation}
    \Omega_{ij} = \frac{Q^2}{2M\sqrt{\omega_i\omega_j}\,r^3_{ij}},
\label{eq:Dipolar coupling strength}
\end{equation}
where the interdipole distances $d_i$ are also uncorrelated random variables distributed uniformly within an interval $[d_0 - \Delta/2, d_0 + \Delta/2]$, the average spacing being $d_0$.
The two dimensionless parameters $W/\omega_0$ and $\Delta/d_0$ hence govern the disorder strength on the dipole frequencies and on the spacings between two neighboring dipoles, respectively.
In our microscopically derived model, the disorder parameters $W/\omega_0$ and $\Delta/d_0$ are restricted, respectively, to values smaller than $2$ and smaller than $2-6a/d_0$, in order to maintain positive resonance frequencies $\omega_i$ and spacings $d_i \geqslant 3a$ which allow us to safely neglect multipolar effects and describe purely dipolar excitations \cite{Park}.

In Eq.~\eqref{eq:Dipolar coupling strength}, $Q<0$ and $M$ are the effective charge and effective mass of a dipole.
For convenience, we recast the two latter parameters in a single one by defining the dipole length scale $a=(Q^2/M\omega_0^2)^{1/3}$.\footnote{As an example, in the case of spherical metallic nanoparticles hosting localized surface plasmons, $Q=-eN_\mathrm{e}$ and $M=m_\mathrm{e}N_\mathrm{e}$, where $N_\mathrm{e}$ is the number of valence electrons of charge $-e<0$ and mass $m_\mathrm{e}$.
The resonance frequency of the plasmon is then the Mie frequency of the nanoparticle $\omega_0 = (N_{\mathrm{e}}e^2/m_{\mathrm{e}}a^3)^{1/2}$, with the dipole length scale $a$ being the nanoparticle radius.
Note that throughout this paper, we use cgs units.}
The average dipolar coupling between two neighboring dipoles is then $\Omega_0 = \overline{\Omega_{i,i+1}} = (\omega_0/2)(a/d_0)^3$, where the bar denotes averaging over the disorder realizations.
In the remainder of our paper, we choose an average spacing $d_0=4a$, so that the dipoles are coupled in the near-field regime, the one most studied both experimentally \cite{Krenn1999,Maier2002,Maier2003,Koenderink2007,Crozier07,Apuzzo2013,Barrow2014} and theoretically \cite{Koenderink,Markel,BrandstetterKunc2016,Downing2017_Retardation}.
Such a small interdipole spacing leads to $\Omega_0/\omega_0=1/128$.

The second term on the r.h.s.\ of the Hamiltonian \eqref{eq:Hamiltonian} describes the photonic degrees of freedom and reads
\begin{equation}
    H_{\mathrm{ph}} = \sum_{\mathbf{k}, \hat{\lambda}_{\mathbf{k}}} \hbar \omega^{\mathrm{ph}}_{\mathbf{k}} {c_{\mathbf{k}}^{\hat{\lambda}_{\mathbf{k}}}}^{\dagger} c_{\mathbf{k}}^{\hat{\lambda}_{\mathbf{k}}}, 
\label{eq:H_ph}
\end{equation}
where the bosonic ladder operator $c_{\mathbf{k}}^{\hat{\lambda}_{\mathbf{k}}}$ (${c_{\mathbf{k}}^{\hat{\lambda}_{\mathbf{k}}}}^{\dagger}$) annihilates (creates) a photon with wavevector $\mathbf{k}$ and transverse polarization $\hat{\lambda}_{\mathbf{k}}$, i.e., $\mathbf{k}\cdot\hat{\lambda}_{\mathbf{k}}=0$.
We consider these photonic degrees of freedom inside a mirror cavity of cuboidal geometry with perfectly conducting walls on all sides, with linear sizes $L_x$, $L_y$, and $L_z$, as sketched in Fig.~\ref{fig:sketch_cavity}.
This amounts to consider hard-wall boundary conditions, and the cavity photon wavevector along the direction $\sigma=x,y,z$ takes the form $k_\sigma=\pi n_\sigma/L_\sigma$, with $n_\sigma \in \mathbb{N}$.
The photonic dispersion relation then reads
\begin{equation}
    \omega^{\mathrm{ph}}_{\mathbf{k}} = \omega^{\mathrm{ph}}_{n_xn_yn_z} = c\, \sqrt{ \left(\frac{\pi n_x}{L_x}\right)^2 + \left(\frac{\pi n_y}{L_y}\right)^2 + \left(\frac{\pi n_z}{L_z}\right)^2 },
\label{eq:Photon dispersion}
\end{equation}
with $c$ being the speed of light in vacuum.

Finally, the last term on the r.h.s.\ of Eq.~\eqref{eq:Hamiltonian} represents the minimal coupling Hamiltonian between dipolar excitations and cavity photons.
In the point-dipole approximation, where we consider a dipole length scale $a$ much smaller than the average inverse wavevector $k_0^{-1} = c/\omega_0$ associated with an individual dipolar mode,\footnote{For this purpose, we choose to fix the value of the dimensionless dipole strength $k_0a$ to $0.1$ in the remainder of this paper.} we have \cite{Craig2012}
\begin{equation}
    H_{\mathrm{dp}\textrm{-}\mathrm{ph}} = \frac{Q}{Mc} \sum_{i=1}^{\mathcal{N}} \mathbf{\Pi}_i \cdot \mathbf{A} (\mathbf{r}_i), 
\label{eq:H_dpph}
\end{equation}
where $\mathbf{r}_i = x\hat{x} + y\hat{y} + z_i\hat{z}$ corresponds to the location of the center of the dipole on site $i$.
By placing the chain at the center of the cavity and separated from the walls in the $z$ direction by a distance $d_0$, we have $x=L_x/2$, $y=L_y/2$ and $z_i=\sum_{l=0}^{i-1}d_l$.
The momentum associated with the $i$th dipole excitation, polarized along the $x$ axis, reads
\begin{equation}
\mathbf{\Pi}_i=\mathrm{i}\sqrt{\frac{M\hbar\omega_i}{2}}\left({b_i}^\dagger - b_i\right)\hat{x},
\label{eq:Momentum}
\end{equation}
while the vector potential quantized in a cuboidal cavity with hard-wall boundary conditions on all sides reads in the Coulomb gauge  \cite{Kakazu1994}
\begin{equation}
    \mathbf{A}(\mathbf{r}_i) = \sum_{\mathbf{k}, \hat\lambda_{\mathbf{k}}} \sqrt{ \frac{2\pi\hbar c^2}{\omega^{\mathrm{ph}}_{\mathbf{k}}} }\left[ \mathbf{f}_{\mathbf{k}}^{\hat{\lambda}_{\mathbf{k}}}(\mathbf{r}_i)c_{\mathbf{k}}^{\hat{\lambda}_{\mathbf{k}}} + {\mathbf{f}_{\mathbf{k}}^{\hat{\lambda}_{\mathbf{k}}}}^{*}(\mathbf{r}_i){c_{\mathbf{k}}^{\hat{\lambda}_{\mathbf{k}}}}^{\dagger}  \right],
\label{eq:Vector potential}
\end{equation}
where the mode functions are
\begin{align}
    \mathbf{f}_{\mathbf{k}}^{\hat{\lambda}_{\mathbf{k}}}(\mathbf{r}_i) &= C
    \begin{pmatrix}
        \cos(k_xx)\sin(k_yy)\sin(k_zz_i)(\hat{x}\cdot\hat{\lambda}_{\mathbf{k}}) \\
        \sin(k_xx)\cos(k_yy)\sin(k_zz_i)(\hat{y}\cdot\hat{\lambda}_{\mathbf{k}}) \\
        \sin(k_xx)\sin(k_yy)\cos(k_zz_i)(\hat{z}\cdot\hat{\lambda}_{\mathbf{k}})
    \end{pmatrix},
\label{eq:Mode functions}
\end{align}
with $C=2/\sqrt{\mathcal{V}}$ if $k_x$, $k_y$, or $k_z=0$, and $C=2\sqrt{2/\mathcal{V}}$ otherwise, $\mathcal{V}=L_xL_yL_z$ being the volume of the cavity.
Note that in our modeling of the light-matter interaction, we do not consider dipole-photon Umklapp processes, the latter being irrelevant in the near-field regime, when the dipole spacings $d_i \ll \lambda_0 = 2\pi/k_0$ \cite{Downing2019,Allard2021}.

In order to simplify the minimal coupling Hamiltonian \eqref{eq:H_dpph}, we use the fact that we consider a number of dipoles $\mathcal{N}\gg1$, 
such that the longitudinal size of the cavity $L_z=z_{\mathcal{N}} + d_0 \gg L_x,L_y$, and choose a cavity aspect ratio $L_y=3L_x$.
We further consider only the resonance between the photonic mode $(n_x,n_y,n_z)  =  (0,1,0)$ and the average dipole resonance frequency, i.e., $\omega^{\mathrm{ph}}_{010} \sim \omega_0$, so that the $y$ and $z$ components of the mode functions \eqref{eq:Mode functions} vanish.
This condition is fulfilled for small cavity heights $3a \lesssim L_x \lesssim 20a$ and it allows us to neglect all of the higher frequency modes such that $\omega^{\mathrm{ph}}_{n_xn_y0} \geqslant 3\omega_0$, which we consider being out of resonance and which, due to the above choice of geometry, are all of the $(n_x,n_y) \neq (0,1)$.
Hence, as justified in greater details in the supplementary materials of Refs.~\cite{Downing2019,Downing2021} (in the case of periodic boundary conditions in the $z$ direction), this resonance condition allows us to keep only the low frequency photonic modes $(0,1,n_z)$, where $n_z$ is chosen in $[1,\mathcal{N}]$, and only one of the two photon polarizations $\hat{\lambda}_{\mathbf{k}}$ is relevant, so that we no longer display the polarization label.

\begin{figure*}[tbh]
 \includegraphics[width=\linewidth]{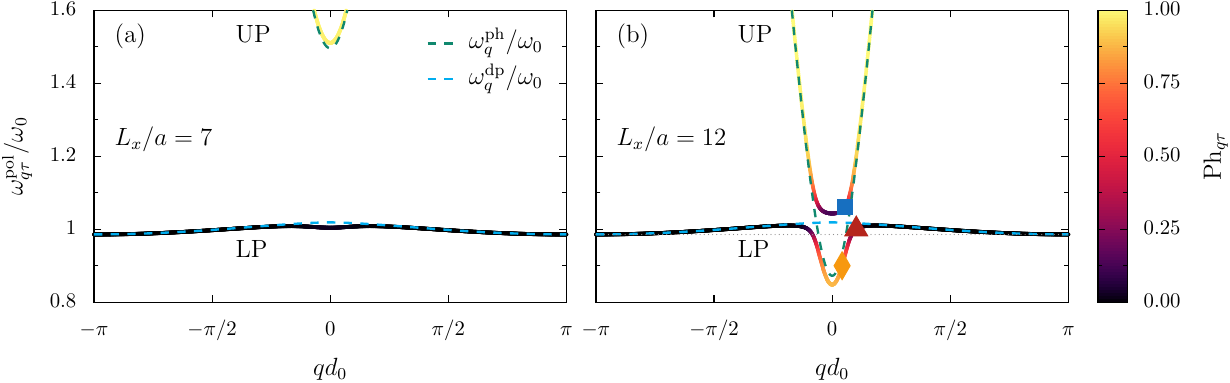}
 \caption{Polaritonic dispersion $\omega_{q\tau}^{\mathrm{pol}}$ [see Eq.~\eqref{eq:Dispersion polariton}] of an infinite, ordered dipole chain for cavity heights $L_x/a = 7$ (left panel) and $L_x/a = 12$ (right panel), in units of the bare frequency $\omega_0$ and as a function of the reduced wavenumber $qd_0$ in the first Brillouin zone. 
 The colormap represents the photonic weight $\mathrm{Ph}_{q\tau}$ of the mode [see Eq.~\eqref{eq:Photonic part fourier}], and the dotted grey line in panel (b) indicates the value of $\omega_{\pm\pi/d_0,-}^{\mathrm{pol}}$. 
 The blue and green dashed lines represent the bare dipolar dispersion $\omega_q^\mathrm{dp}$ [see Eq.~\eqref{eq: Dipole dispersion RWA Fourier}] and the bare photonic one $\omega_q^\mathrm{ph}$ [see Eq.~\eqref{eq:Photon dispersion approximated Fourier}], respectively. The orange, red, and blue symbols point out eigenstates that will be studied in detail in the sequel.
 As in the remaining of this paper, the cavity aspect ratio is fixed to $L_y = 3L_x$, the interdipole spacing is $d_0=4a$, and the dimensionless dipole strength $k_0a=0.1$.}
 \label{fig:Dispersion Fourier}
\end{figure*}

Within the above-discussed hypotheses and using in addition the rotating wave approximation, which we discuss in Appendix~\ref{sec:RWA}, the Hamiltonian \eqref{eq:Hamiltonian} becomes
\begin{align}
    \tilde{H} =&\, H_\mathrm{dp} + \sum_{n_z=1}^{\mathcal{N}} \hbar\omega^{\mathrm{ph}}_{n_z}{c^\dagger_{n_z}} c^{\phantom{\dagger}}_{n_z} \nonumber \\
    &\, + \mathrm{i}\hbar\sum_{i=1}^{\mathcal{N}}\sum_{n_z=1}^{\mathcal{N}}\xi_{in_z}\left(b_i^\dagger c^{\phantom{\dagger}}_{n_z} - b^{\phantom{\dagger}}_i c_{n_z}^\dagger \right),
\label{eq:Hamiltonian approximated}
\end{align}
where the photonic dispersion \eqref{eq:Photon dispersion} simplifies to 
\begin{equation}
    \omega^{\mathrm{ph}}_{n_z} = c\,\sqrt{ \left( \frac{\pi}{L_y} \right)^2 + \left( \frac{\pi n_z}{L_z} \right)^2 },
\label{eq:Photon dispersion approximated}
\end{equation}
the light-matter coupling writes
\begin{equation}
    \xi_{in_z} = \omega_0\sqrt{ \frac{4\pi a^3\omega_i}{\mathcal{V}\omega^{\mathrm{ph}}_{n_z}} }\sin\left(\frac{\pi n_z}{L_z}z_i \right),
\label{eq:Light-Matter coupling}
\end{equation}
and with the dipolar Hamiltonian $H_\mathrm{dp}$ being given in Eq.~\eqref{eq:H_dp}.

In the following, we will first study the properties of the spectrum and eigenstates of the approximate Hamiltonian \eqref{eq:Hamiltonian approximated}, and then study the propagation along the chain in a driven-dissipative transport scenario, considering losses both in the dipoles and in the cavity mirrors.
Furthermore, we emphasize that even if we model the system using a fully quantum theory, the basic ingredients, namely dipolar excitations interacting with confined electromagnetic modes, are of purely classical nature and no quantum effects are considered here. 
Hence, the system could also be modeled using a classical formalism relying on Maxwell's equations, which is however, in our view, less transparent and less straightforward to implement  \cite{Koenderink,Markel,Mann2020,Fernique2020,Allard2021,Mann2022}.

\section{Eigenspectrum analysis and localization properties}
\label{sec:Spectrum}

\subsection{Ordered chain in the thermodynamic limit}
\label{subsec:Ordered}

To gain insight on the model and for clarity, we begin by summarizing and further discussing results recently obtained in Ref.~\cite{Downing2021} for the simple case of an ordered chain in the thermodynamic limit, which can be solved analytically.
Within the framework developed in Sec.~\ref{sec:Model}, this implies to consider the disorder parameters $W=0$ and $\Delta=0$, such that $\omega_i = \omega_0$ and $d_i = d_0$.
By going to the limit $\mathcal{N} \xrightarrow{}\infty$, we can use periodic boundary conditions for both the dipole chain and the cavity in the $z$ direction.
This allows one to move into wavevector space and to Fourier diagonalize the periodic boundary condition version of the Hamiltonian \eqref{eq:Hamiltonian approximated}.
For consistency, we give in Appendix \ref{sec:Diagonalization} some details about the Fourier diagonalization, which can also be found in Ref.~\cite{Downing2021}.

The result of the diagonalization procedure with periodic boundary condition is shown is Fig.~\ref{fig:Dispersion Fourier}, where we plot the polaritonic dispersion $\omega_{q\tau}^\mathrm{pol}$ [see Eq.~\eqref{eq:Dispersion polariton}] in the first Brillouin zone (BZ), for cavity heights $L_x=7a$ [Fig.~\ref{fig:Dispersion Fourier}(a)] and $L_x=12a$ [Fig.~\ref{fig:Dispersion Fourier}(b)].
The colormap represents the photonic part $\mathrm{Ph}_{q\tau}$ of each eigenmodes, given in Eq.~\eqref{eq:Photonic part fourier}, and we hereafter refer to the high (low) frequency band, labeled by the index $\tau=+$ ($-$), as the upper (lower) polariton UP (LP).
We also plot by blue and green dashed lines the bare dipolar $\omega_q^\mathrm{dp}$ and bare photonic $\omega_q^\mathrm{ph}$ dispersions, given respectively in Eqs.~\eqref{eq: Dipole dispersion RWA Fourier} and \eqref{eq:Photon dispersion approximated Fourier}.

We first notice that increasing the cavity height $L_x$ (and hence $L_y$, since the aspect ratio is fixed to $L_y=3L_x$) leads to an increasing renormalization of the bare dispersions $\omega_q^\mathrm{dp}$ and $\omega_q^\mathrm{ph}$, with an increasingly pronounced avoided crossing, typical of the strong-coupling regime.
When $L_x$ becomes sufficiently large, namely for $L_x \geqslant \omega_0\pi L_x/\omega_{q=0}^\mathrm{dp}k_0L_y$ which corresponds to $L_x \simeq 10.3a$ with the parameters used in Fig.~\ref{fig:Dispersion Fourier}, the light-matter detuning $\Delta_q = (\omega_q^{\mathrm{ph}} - \omega_q^{\mathrm{dp}})/2$ can be smaller than $0$, meaning that there is a crossing between the bare dipolar and photonic dispersions at a given wavenumber $q_\mathrm{res}$.
From this point on, we can thus identify $\xi_{q_{\mathrm{res}}}$ [see Eq.~\eqref{eq:Light-matter coupling fourier}] with half of the Rabi splitting frequency.
In the case of $L_x/a = 12$ shown in Fig.~\ref{fig:Dispersion Fourier}(b), the Rabi splitting frequency $\Omega_{\mathrm{RS}} \simeq 0.12 \omega_0$.

We note that increasing the cavity height $L_x$ reduces the light-matter coupling $\xi_q$ in units of $\omega_0$, as can be seen from Eq.~\eqref{eq:Light-matter coupling fourier}.
This can be understood by noting that the light-matter detuning $\Delta_q$ is reduced when the cavity height increases, due to the cavity size dependence of the bare photonic dispersion $\omega_q^\mathrm{ph}$ [see Eq.~\eqref{eq:Photon dispersion approximated Fourier}].
This behavior will be reversed in the limit of large cavity size $L_x \gg a$, where our low-frequency approximation \eqref{eq:Hamiltonian approximated} breaks down, and where one recovers the regime of a dipolar chain coupled to vacuum electromagnetic modes only, studied, e.g., in Refs.~\cite{Downing2017_Retardation,Allard2021}.

From the colormap of Fig.~\ref{fig:Dispersion Fourier}, one can see that the hybridization between light and matter degrees of freedom is enhanced by increasing the cavity height $L_x$, in agreement with the entrance in the strong-coupling regime.
The most hybridized states are the ones near the center of the BZ, for which the detuning $\Delta_q$ is around and below $0$ [see expression \eqref{eq:Photonic part fourier}].
In the following, we will refer to these highly hybridized states as \emph{polaritons}.
On the other hand, states that are apart from the center of the BZ, for which  the detuning $\Delta_q$ is positive, remain almost unhybridized.
The states in the LP branch that remain almost fully dipolar, namely with a photonic weight $\mathrm{Ph}_{q\tau} \lesssim 0.1$, will be referred to as \emph{dark states}\footnote{In contrast to single cavity mode models such as the Tavis-Cummings one, in a realistic multimode model there is no unambiguous definition of dark and polaritonic states \cite{Ribeiro2022}, all the eigenstates being hybridized.} (visible in black in Fig.~\ref{fig:Dispersion Fourier}). 
Unlike polaritons, these dark states all have an eigenfrequency very close to that of the isolated dipole frequency $\omega_0$.

Finally, we indicate by a grey dotted line in Fig.~\ref{fig:Dispersion Fourier}(b) the eigenfrequency $\omega_{\pm\pi/d_0,-}^{\mathrm{pol}}$ of the darkest, least coupled eigenmode of the LP branch, sitting at the edge of the BZ.
This highlights the fact that due to the dipolar coupling between the emitters, which leads to the collective dispersion relation given in Eq.~\eqref{eq: Dipole dispersion RWA Fourier}, there are both LP eigenstates with a higher and lower eigenfrequency than that of the least coupled one.
In the next subsection, by studying the localization properties of the disordered chain we will show that this at first sight trivial information becomes important, since the latter eigenstates presenting an eigenfrequency lower than $\omega_{\pm\pi/d_0,-}^{\mathrm{pol}}$ will be found to be particularly robust against disorder, while the ones above the dotted line constitute what will later be called the dark state band.

\subsection{Finite disordered chain}
\label{subsec:Disordered}
 
We now move to a numerical study of the eigenstates of the finite disordered dipole chain, considering the real space Hamiltonian \eqref{eq:Hamiltonian approximated}, that can be written in a $2\mathcal{N}\times2\mathcal{N}$ matrix form using the basis vector $\mathbf{\varphi}^\dagger = (b_1^\dagger,\dots,b_\mathcal{N}^\dagger,c_1^\dagger,\dots,c_\mathcal{N}^\dagger)$.
In this context, the dipolar and photonic parts of a given eigenstate $n \in [1,2\mathcal{N}]$, that is, the real space counterparts of $\mathrm{D}_{q\tau}$ and $\mathrm{Ph}_{q\tau}$ given in Eq.~\eqref{eq:Dipolar and photonic part fourier}, are given by
\begin{equation}
    \mathrm{D}(n) = \sum_{i=1}^{\mathcal{N}}|\Psi_i(n)|^2\;\;\;\;\;\textrm{and}\;\;\;\;\; \mathrm{Ph}(n) = 1 - \mathrm{D}(n),
\label{eq:Photonic part real}
\end{equation}
where the $2\mathcal{N}$-component normalized eigenvector $\Psi(n) = (\Psi_1(n),\dots,\Psi_{2\mathcal{N}}(n))$, the first $\mathcal{N}$ entries corresponding to the dipolar subspace.

To characterize the localization of an eigenstate $n$, we use the participation ratio (PR) defined as \cite{Bell1970,Thouless1974}
\begin{equation}
    \mathrm{PR}(n) = \frac{ \left(  \sum_{i=1}^{\mathcal{N}}|\Psi_i(n)|^2  \right)^2  }{ \sum_{i=1}^{\mathcal{N}}|\Psi_i(n)|^4   },
\label{eq:Participation Ratio}
\end{equation}
where the summations are taken over the dipolar subspace only.
The quantity \eqref{eq:Participation Ratio} gives information about the typical number of sites $i$ occupied by an eigenstate $n$. In 1d systems it is then proportional to the localization length.
Extended states are characterized by a PR scaling with the total number of sites $\mathcal{N}$, while the PR of localized states is size-independent.
We insist on the fact that in this paper, when we refer to an extended or delocalized state, we mean \emph{delocalized at the scale of the system size}, namely, that the localization length is larger that the system.
Indeed, no genuine finite size scaling argument has been performed to confirm that such states are still extended at the thermodynamic limit, as states that can be found, e.g., in three-dimensional disordered systems featuring an Anderson transition.
A discussion about this possible ambiguity, including the scaling of the PR with the system size for $\mathcal{N} \in [100,5000]$, is proposed in Appendix~\ref{sec:multifractality}.

Here and hereafter, we fix the interdipole spacing disorder $\Delta$ to $0$ and focus only on the effects of the frequency disorder $W$.
Due to the fact that the individual frequencies $\omega_i$ appear both on- and off-diagonal in the dipolar Hamiltonian [see Eq.~\eqref{eq:H_dp}], the frequency disorder affects the spectrum more than the interdipole spacing one $\Delta$, which is purely off-diagonal.
All the results shown here are then qualitatively the same with a spacing disorder $\Delta \neq 0$.
The particular case of off-diagonal disorder only which, remarkably, yields to the Dyson singularity in the density of states \cite{Dyson1953}, would however not lead to important disorder-induced effects as we observe here.
It is thus beyond the scope of this study.
\subsubsection{Disorder-induced mixing of dipolar and photonic weights}

\begin{figure*}[tb]
    \includegraphics[width=\columnwidth]{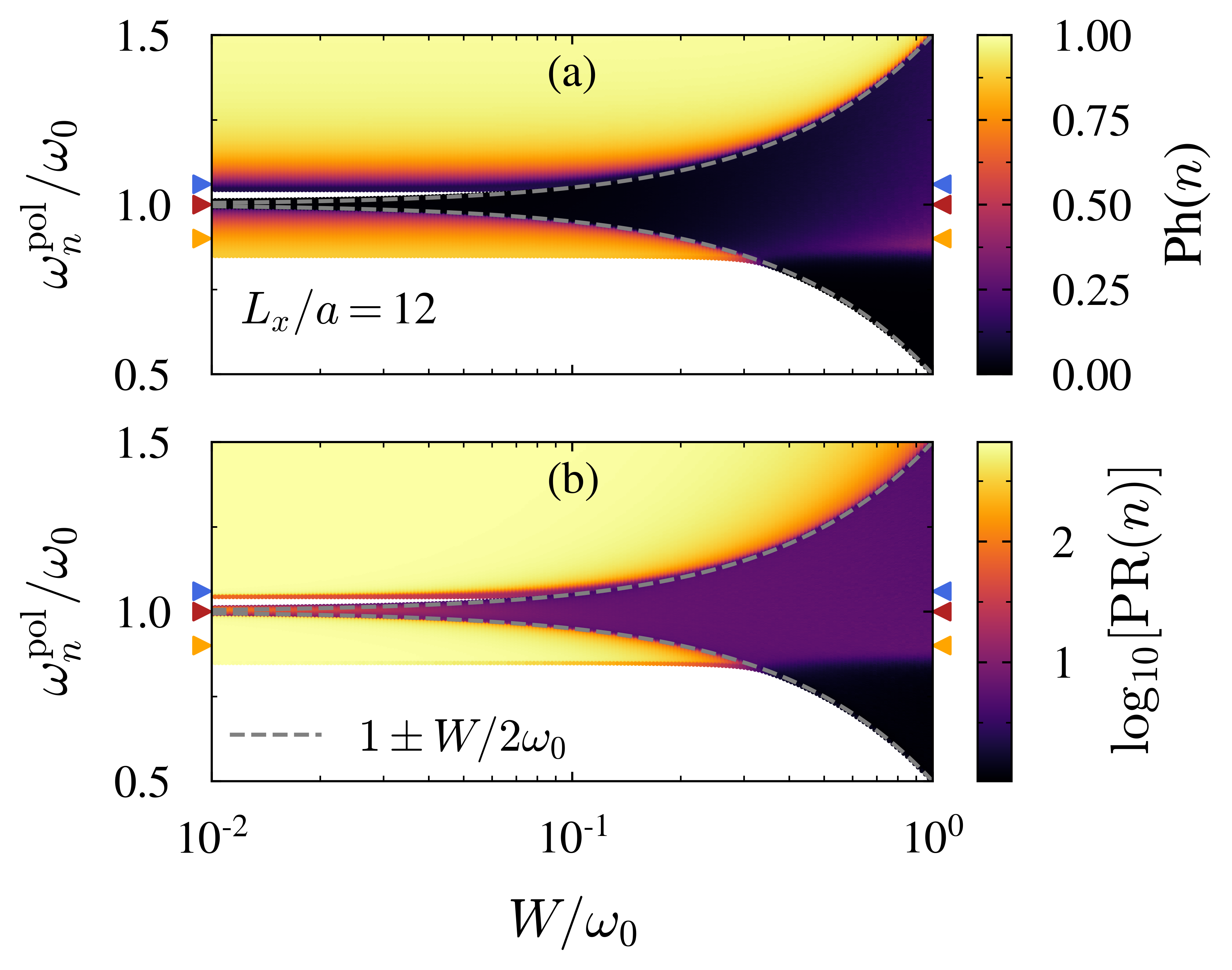}
    \hspace{0.48cm}
    \includegraphics[width=\columnwidth]{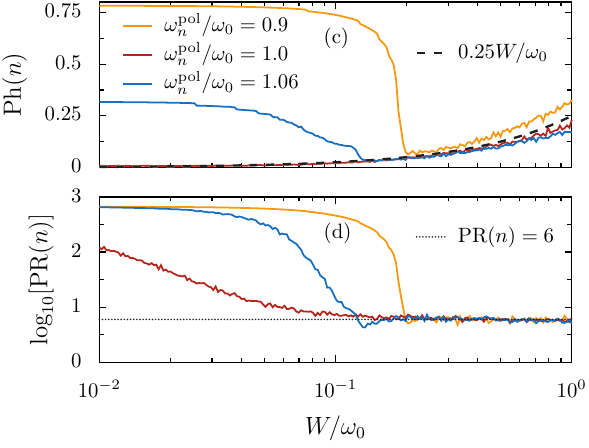}
    \caption{(a)-(b) Colormaps of the photonic weight $\mathrm{Ph}(n)$ and of the participation ratio $\mathrm{PR}(n)$ as a function of both the eigenfrequencies $\omega_n^\mathrm{pol}$ and disorder strength $W$ in units of the average dipole frequency $\omega_0$.
    (c)-(d) Horizontal cuts of panels (a)-(b) at the eigenfrequencies $\omega_n^{\mathrm{pol}}/\omega_0=0.9$, $1.0$, and $1.06$, pointed out in panels (a)-(b) by orange, red, and blue triangles, respectively.
    The cavity height is $L_x/a=12$, the number of dipole is $\mathcal{N}=1000$, and the data have been averaged over $100$ disorder realizations.}
\label{fig:Freq, PR and Ph function of W}
\end{figure*}

To begin our study of a disordered chain of dipoles in a photonic cavity, we place ourselves in the strong-coupling regime, namely with a cavity height $L_x/a = 12$ [see Fig.~\ref{fig:Dispersion Fourier}(b) for the ordered counterpart], and we consider a chain of $\mathcal{N} = 1000$ dipoles.
We compute the average of the photonic part $\mathrm{Ph}(n)$ and of the participation ratio $\mathrm{PR}(n)$ of the eigenmodes $n$ of the system over $100$ disorder realizations, for increasing frequency disorder strength $W$.
The result is shown in Figs.~\ref{fig:Freq, PR and Ph function of W}(a) and \ref{fig:Freq, PR and Ph function of W}(b), as a function of both the eigenfrequencies $\omega_n^{\mathrm{pol}}$ which are ordered in ascending order, and the disorder strength $W$.

We first discuss the spectrum for weak disorder strengths $W/\omega_0 \simeq 10^{-2}$, which could correspond typically to experimental uncertainties obtained in the fabrication of plasmonic nanoparticles, namely inhomogeneities in their sizes \cite{Koenderink2007,Apuzzo2013,Mueller2020}, resulting in different resonance frequencies $\omega_i$ \cite{Soennichsen2002PRL,Soennichsen2002,Berciaud2005}. In Fig.~\ref{fig:Freq, PR and Ph function of W}(a), we recognize the same behavior as with an ordered chain (see Fig.~\ref{fig:Dispersion Fourier}), namely that the bottom of the LP branch is mainly photonic, visible in orange, whereas the top is almost purely dipolar, visible in black around $\omega_n^{\mathrm{pol}}/\omega_0 = 1.0$.
Furthermore, the bottom of the UP branch is mainly dipolar, while the rest of the band goes from predominantly photonic to almost purely photonic states.
By comparing these results to the PR in Fig.~\ref{fig:Freq, PR and Ph function of W}(b), we observe that mainly photonic states have a very high PR, i.e., they are delocalized along the chain, while the almost purely dipolar dark states show a lower PR, hence being localized on a small number of sites.
This expected behavior shows the cavity-protection effect, namely that polaritonic states are more robust against disorder than purely dipolar states \cite{Houdre1996}.
We note that mainly dipolar states with a photonic weight $\mathrm{Ph}(n) \simeq 0.25$, visible in purple in Fig.~\ref{fig:Freq, PR and Ph function of W}(a), already benefit from this effect.

By now increasing disorder, we observe in Figs.~\ref{fig:Freq, PR and Ph function of W}(a) and \ref{fig:Freq, PR and Ph function of W}(b), as expected, an increase of the bandwidth of the polariton bands.
This amounts to the closing of the bandgap such that we cannot unambiguously distinguish between LP and UP branches anymore, so that the system is no longer strictly in the strong-coupling regime.
Interestingly, the growing portion of the spectrum is the one containing the dark states, namely the top of the LP branch, which in this sense behaves as an effective band.
This observation agrees with Ref.~\cite{Yang2022}, where similar behavior has been encountered.
All the states outside this effective dark state band profit from the cavity-protection effect.
We note that the dark state band is bounded from below by the least coupled, darkest mode, which, in the ordered chain picture, was marked as the dotted grey line in Fig.~\ref{fig:Dispersion Fourier}.
As shown by the grey dashed lines in Figs.~\ref{fig:Freq, PR and Ph function of W}(a) and \ref{fig:Freq, PR and Ph function of W}(b), the rate of expansion of the dark state band around $\omega_0$ is $W/2$, that is, half the width of the rectangular distribution in which the individual dipole frequencies $\omega_i$ are randomly chosen.
This is due to the fact that dark states have eigenfrequencies which are almost unchanged by cavity photons, so that they remain of the order of $\omega_i$.

\begin{figure*}[tb]
    \includegraphics[width=\linewidth]{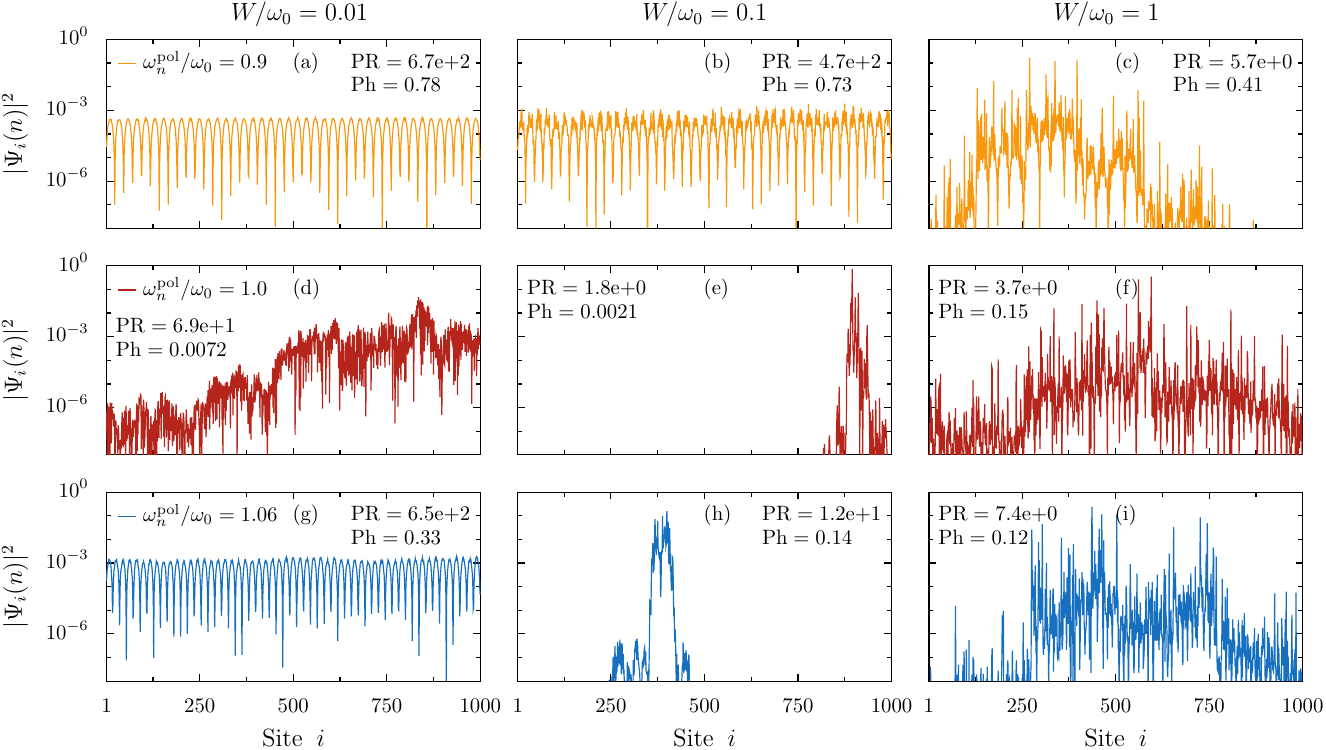}
    \caption{Probability density $|\Psi_i(n)|^2$ along the sites $i$ of a chain of $\mathcal{N}=1000$ dipoles, for different eigenfrequencies  $\omega_n^\mathrm{pol}/\omega_0 = 0.9$, $1.0$, and $1.06$ (rows), and different disorder strength $W/\omega_0=0.01$, $0.1$, and $1$ (columns).
    In the figure, data are not averaged over disorder realizations and the cavity height $L_x/a=12$.}
\label{fig:Probability density for different disorder}
\end{figure*}

When this broadening of the eigenfrequencies approximately reaches the bottom of the LP branch, that is, when the disorder strength $W \gtrsim 0.3$, fully dipolar eigenstates which are totally localized with a $\mathrm{PR}(n) \simeq 1$ appear in the spectrum [see the black region at the bottom right of Figs.~\ref{fig:Freq, PR and Ph function of W}(a) and \ref{fig:Freq, PR and Ph function of W}(b)].
In that respect and as detailed in the next subsection, for a given large enough disorder strength $W \gtrsim 0.3$ the model hosts three distinct phases as can be seen from Figs.~\ref{fig:Freq, PR and Ph function of W}(a) and \ref{fig:Freq, PR and Ph function of W}(b): an exponentially localized, dark state phase composed of the eigenfrequencies $\omega_n^{\mathrm{pol}} \lesssim \omega_{q=0}^\mathrm{ph}$ (black region), a phase of mostly photonic states, not yet affected by disorder due to their eigenfrequencies $\omega_n^{\mathrm{pol}} \gtrsim \omega_0 + W/2$ (yellow-orange region), and a phase of mainly dipolar polaritons with an intermediate PR, made of states in the dark state band with eigenfrequencies $\omega_{q=0}^\mathrm{ph} \lesssim \omega_n^{\mathrm{pol}} \lesssim \omega_0 + W/2$ (purple region).

The latter intermediate phase presents the most interesting properties.
We highlight them in Figs.~\ref{fig:Freq, PR and Ph function of W}(c) and \ref{fig:Freq, PR and Ph function of W}(d), by showing horizontal cuts of panels (a) and (b), for the eigenmodes corresponding to the eigenfrequencies $\omega_n^\mathrm{pol}/\omega_0=0.9$, $1.0$, and $1.06$, plotted as orange, red, and blue solid lines, respectively.
These eigenstates are also indicated, respectively, by an orange diamond, a red triangle, and a blue square in the ordered bandstructure of Fig.~\ref{fig:Dispersion Fourier}(b).
At weak disorder strength, they correspond, respectively, to a mainly photonic polariton, a dark state, and a mainly dipolar polariton.
Crucially, we observe in Fig.~\ref{fig:Freq, PR and Ph function of W}(c) that by increasing the disorder strength, the dark state at $\omega_n^\mathrm{pol}/\omega_0=1.0$ (red line) has a photonic weight which increases linearly with the frequency disorder, at a rate around $0.25W/\omega_0$, as exemplified by the black dashed line.
This disorder-induced gain of photonic weight leads to the fact that no more dark states are present at the frequency where they were present without disorder, i.e., around $\omega_0$, since disorder hybridizes all of them to polaritons.
Furthermore, the mainly photonic polariton at $\omega_n^\mathrm{pol}/\omega_0=0.9$ (orange line) changes drastically in nature when the disorder strength is increased, becoming a mainly dipolar polariton when its frequency enters the dark state band, that is when $W \gtrsim 2|\omega_n^{\mathrm{pol}} - \omega_0|=0.2$.
Once in the dark state band, its photonic weight also follows a linear increase with disorder strength.
Finally, the same mechanism occurs for the mainly dipolar polariton at $\omega_n^\mathrm{pol}/\omega_0=1.06$ (blue line).
The corresponding PR values in Fig.~\ref{fig:Freq, PR and Ph function of W}(d) reveal that once in the dark state band, the states show an intermediate value of the PR, here $\mathrm{PR}(n) \simeq 6$, which, as opposed to the photonic weight, remains constant when the disorder strength increases.
Importantly, this means that the PR of these states does not fall to $1$ with strong disorder strength, in contrast to what is usually the case in 1d disordered systems.
This can be understood from the fact that the disorder-induced hybridization of the dark states into polaritons allows them to inherit the polariton robustness against localization, that is, the cavity-protection effect.

In order to better understand the mixing of dipolar and photonic weights induced by disorder, we display in Fig.~\ref{fig:Probability density for different disorder} the probability density $|\Psi_i(n)|^2$ along the sites $i$ of the chain for the same three eigenstates, for which the same color code is used.
It should be noted, however, that the results of Fig.~\ref{fig:Probability density for different disorder} correspond to a given disorder realization. 
Increasing values of disorder strength $W/\omega_0=0.01, 0.1$, and $1$ are considered from left to right panels.

In Figs.~\ref{fig:Probability density for different disorder}(a)-\ref{fig:Probability density for different disorder}(c), we present our results for the mainly photonic polariton with eigenfrequency $\omega_n^\mathrm{pol}/\omega_0 = 0.9$.
By increasing the disorder strength, the state goes from delocalized and mostly photonic in panels (a) and (b), to localized on multiple nonadjacent sites with a very small PR in panel (c).
In the following we term these states ``semilocalized", as recently proposed by Botzung \textit{et al.}\ in Ref.~\cite{Botzung2020}, where they unveiled similar phenomena in a disordered Tavis-Cummings model.
As studied in Refs.~\cite{Botzung2020,Chavez21,Dubail22}, semilocalized states can be seen as localized states, that is, an exponential peak with a size-independent PR, with long tails that have increased so that they are of the same order of magnitude as the original peak.
In Appendix~\ref{sec:multifractality}, we discuss the nature of the semilocalized states present in our model, and notably the differences with the ones found in Refs.~\cite{Botzung2020,Chavez21,Dubail22}.
In our model, these semilocalized states correspond to the purple region in Figs.~\ref{fig:Freq, PR and Ph function of W}(a) and \ref{fig:Freq, PR and Ph function of W}(b), i.e., they belong to the dark state band with eigenfrequencies $\omega_{q=0}^\mathrm{ph} \lesssim \omega_n^{\mathrm{pol}} \lesssim \omega_0 + W/2$.
They present a constant value of PR of the order of $10$, as well as a photonic weight increasing 
linearly with the disorder strength.

In the second row of Fig.~\ref{fig:Probability density for different disorder}, we show the same quantities for a dark state with eigenfrequency $\omega_n^\mathrm{pol}/\omega_0 = 1.0$.
As one can see from panel (d), this fully dipolar state already becomes affected by a weak disorder strength $W/\omega_0=0.01$, but the exponential localization occurs only at a stronger disorder strength $W/\omega_0=0.1$, visible in panel (e).
We note that in addition to the exponential peak, exponentially localized states also include algebraic tails with very small amplitude.
By increasing further the disorder strength to $W/\omega_0=1$ in panel (f), the state acquires a photonic weight and becomes semilocalized.
The same mechanism is visible in the third row of Fig.~\ref{fig:Probability density for different disorder} [panels (e)-(g)] for a mainly dipolar polariton with eigenfrequency $\omega_n^\mathrm{pol}/\omega_0 = 1.06$, which, due to its polaritonic nature, remains unaffected at weak disorder strength [see Fig.~\ref{fig:Probability density for different disorder}(g)].

\subsubsection{Influence of the light-matter coupling}

\begin{figure*}[tb]
    \includegraphics[width=\columnwidth]{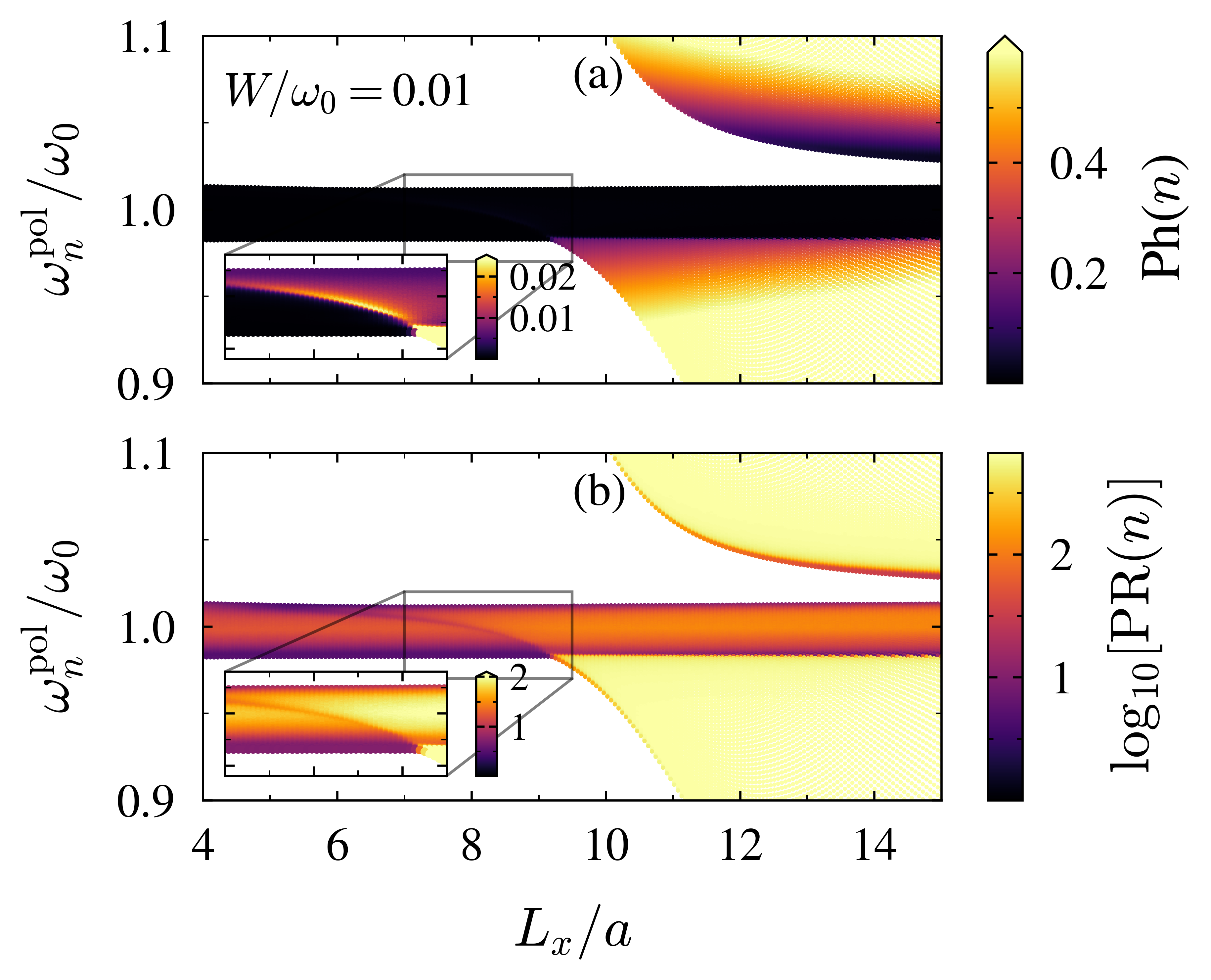}
    \includegraphics[width=\columnwidth]{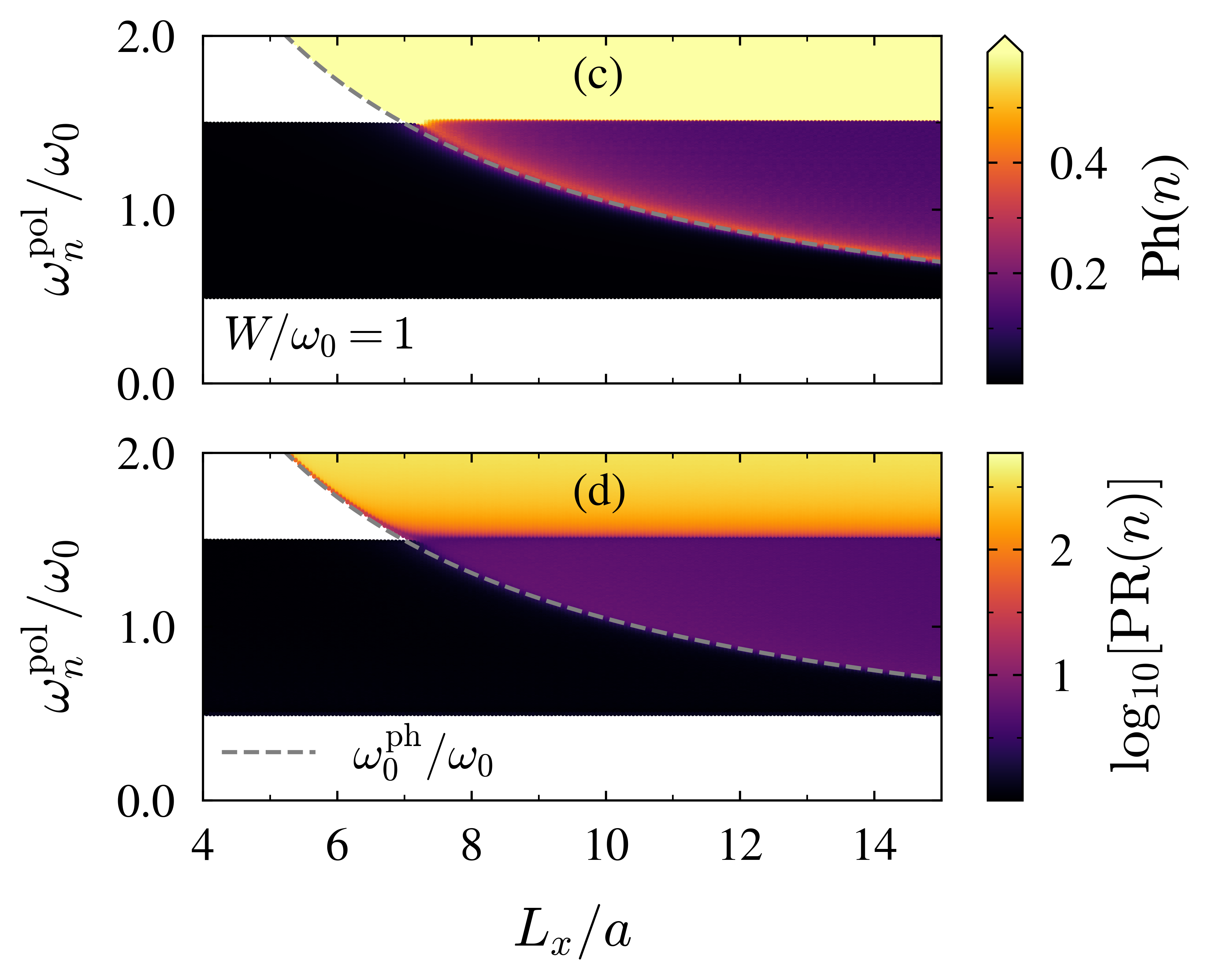}
    \caption{(a),(c) Photonic weight $\mathrm{Ph}(n)$ and (b),(d) participation ratio $\mathrm{PR}(n)$ as a function of both the reduced eigenfrequencies $\omega_n^{\mathrm{pol}}/\omega_0$ and cavity height $L_x/a$, for disorder strengths $W/\omega_0 = 0.01$ [left panels (a)-(b)] and $W/\omega_0 = 1$ [right panels (c)-(d)]. The data have been averaged over $100$ disorder realizations.}
\label{fig:Frequency function of Lx}
\end{figure*}

Before going to transport computations, here we conclude the eigenspectrum analysis by studying the influence of the light-matter coupling on the eigenstate properties discussed previously.
For that purpose, we show in Fig.~\ref{fig:Frequency function of Lx} the photonic weight [Eq.~\eqref{eq:Photonic part real}] and the PR [Eq.~\eqref{eq:Participation Ratio}] of the eigenmodes as a function of both the eigenfrequencies ordered in ascending order, and the cavity height $L_x$ that controls the light-matter coupling.

In the left panels, Figs.~\ref{fig:Frequency function of Lx}(a)-\ref{fig:Frequency function of Lx}(b), a weak disorder strength $W/\omega_0 = 0.01$ is considered.
As can be seen from panel (a), in the weak-coupling regime ($L_x/a \lesssim 9$) the eigenstates are almost not hybridized and the LP branch corresponds only to dark states.
The corresponding PR in panel (b) shows that these dark states are already localized, especially the modes near the band edges, whereas the ones around the middle of the band present a higher PR.
This behavior of sharply localized band edges is usual in all 1d Anderson-like disordered systems \cite{Mirlin_RevModPhys}.
In the insets of Figs.~\ref{fig:Frequency function of Lx}(a) and \ref{fig:Frequency function of Lx}(b), we show a zoom around the modes which interact the most with cavity photons, namely the ones at the center of the BZ in the ordered chain picture (around $q=0$ in Fig.~\ref{fig:Dispersion Fourier}). 
The inset of panel (a) shows that theses modes have a higher photonic weight than the surrounding ones, being the most hybridized states. 
However, the inset of panel (b) demonstrates that these states have a smaller PR than the surrounding ones, hence being more localized, despite their larger photonic weight. 
This counterintuitive behavior highlights the particularly nontrivial interplay between localization and light-matter hybridization.

At larger cavity height ($L_x/a \gtrsim 9$), we see in Figs.~\ref{fig:Frequency function of Lx}(a)-\ref{fig:Frequency function of Lx}(b) the emergence of polaritons at the bottom of the LP branch, with a high PR value.
The existence of these eigenstates is permitted by the strong enough band modification due to the avoided-crossing, which allows eigenfrequencies smaller than the one of the darkest mode [which was marked as the grey dotted line in Fig.~\ref{fig:Dispersion Fourier}(b)].\footnote{This unusual transition, occurring here at a cavity height $L_x/a \simeq 9.1$, originates solely from the $1/r^3$ quasistatic dipole-dipole coupling that leads to a dipolar band which is not flat. Indeed, a flat dipolar dispersion, as considered in usual strong-coupling toy models, would lead to eigenfrequencies that are smaller than the most uncoupled one for any nonzero value of the light-matter coupling.}
In contrast, the top of the LP branch, that we previously coined the dark state band, remains composed of almost fully dipolar states [see panel (a)].
Looking at the corresponding PR in panel (b), one sees that the modes near the band edges of this dark state band are more localized than the ones in the middle, confirming the behavior of an effective band.

In the right panels, Figs.~\ref{fig:Frequency function of Lx}(c)-\ref{fig:Frequency function of Lx}(d), we consider now the case of a larger disorder strength $W/\omega_0 = 1$.
As we increase the cavity height $L_x/a$, we observe an increase of the share of semilocalized states, visible in purple both in Figs.~\ref{fig:Freq, PR and Ph function of W}(a)-\ref{fig:Freq, PR and Ph function of W}(b) and \ref{fig:Frequency function of Lx}(c)-\ref{fig:Frequency function of Lx}(d).
The frequency range occupied by these semilocalized states matches the part of the dark state band overlapping with the bare photonic band, thus being bounded from below by the lowest bare photonic state $\omega_{q=0}^{\mathrm{ph}}$, which we represent as a function of $L_x/a$ as a grey dashed line.
This demonstrates that in our model, semilocalization can be understood as eigenstates being mixed to photonic states by the increase of the bandwidth induced by the frequency disorder.
This permits new coupling between dipolar and photonic degrees of freedom.
States with an eigenfrequency $\omega_n^{\mathrm{pol}} \lesssim \omega_{q=0}^{\mathrm{ph}}$, visible in black in Figs.~\ref{fig:Frequency function of Lx}(c)-\ref{fig:Frequency function of Lx}(d), are the fully dipolar exponentially localized states, namely, dark states subject to Anderson localization.
Due to their eigenfrequencies being lower than the ones of the photonic band, they cannot benefit from the disorder-induced mixing between dipolar and photonic degrees of freedom. 
On the other hand, states not already reached by the dark state band, with $\omega_n^{\mathrm{pol}} \gtrsim \omega_0 + W/2$, remain mainly photonic and delocalized [see the orange/yellow regions in Figs.~\ref{fig:Frequency function of Lx}(c)-\ref{fig:Frequency function of Lx}(d)], confirming the previous analysis of Figs.~\ref{fig:Freq, PR and Ph function of W}(a)-\ref{fig:Freq, PR and Ph function of W}(b).
Finally, we note that with such a large value of the disorder strength ($W/\omega_0 = 1$), we are able to see semilocalized states already for a small cavity height $L_x/a \simeq 7$, a value from which we considered the system being in the weak-coupling regime without disorder.
The transition from dark to semilocalized states when increasing the disorder strength is smooth, but with the above explanation, we understand that for a given value of $W$, semilocalized states would be present only if $2|\omega_{q=0}^{\mathrm{ph}} - \omega_0| \lesssim W$, the light-matter coupling competing with disorder \cite{Botzung2020}.

To conclude this section, by studying the eigenspectrum of our system, we have shown that the bandwidth increase led by frequency disorder tends to incorporate photonic states to the dark state band, leading to states that increase their photonic weight and become semilocalized.
Almost fully dipolar dark states, such as the ones shown in panels (d)-(f) of Fig.~\ref{fig:Probability density for different disorder}, can thus undergo first a transition from being delocalized along the chain to being exponentially localized on a few sites, and then another transition to being semilocalized and turning into hybridized polaritonic states.
As we will show in the next section by performing transport simulations in a driven-dissipative scenario, the latter hybridized states inherit long-range polaritonic transport properties.
This remarkable phenomenon thus leads to a disorder-enhanced propagation along the chain, that is, to the decrease and then the re-increase of the long-range transport properties for increasing disorder.

\section{Driven-dissipative transport scenario}
\label{sec:Transport}

We now move to transport simulations in order to elucidate the fate of the cavity-protection effect, as well as the disorder-induced mixing between dipolar and photonic degrees of freedom and semilocalization in the propagation characteristics along the chain.
Especially, we here take into account inevitable losses both in the dipoles and in the cavity mirrors.
For that purpose, we consider a driven-dissipative transport scenario by adding to the polaritonic Hamiltonian $\tilde{H}$ given in Eq.~\eqref{eq:Hamiltonian approximated} a driving term 
\begin{equation}
    H_{\mathrm{drive}}(t) = \hbar\Omega_\mathrm{R} f\!\left(t\right) \left( b_1^\dagger + b_1^{\phantom{\dagger}} \right),
    \label{eq:H_drive}
\end{equation}
which corresponds to a transversely polarized electric field with amplitude $E_0$ acting on the first dipole, with the Rabi frequency $\Omega_\mathrm{R} = E_0 \sqrt{  {Q^2}/{2M\hbar\omega_0}  }$,
and where $f(t)$ is a time-dependent function that depends on the driving frequency $\omega_\mathrm{d}$.
In the following, we consider a drive $f\left(t\right) = \mathrm{e}^{\mathrm{i}\omega_\mathrm{d} t}$, corresponding to the first site of the chain being continuously illuminated by a monochromatic electric field.

We assume that the propagation dynamics can be approximated by the Lindblad master equation for the density matrix
\begin{align}
    \dot{\rho} =&\; \frac{\mathrm{i}}{\hbar}\left[ \rho, \tilde{H}+H{_\mathrm{drive}}(t) \right] 
    - \sum_{i} \frac{\gamma_i}{2} \left(  \left\{ b_i^\dagger b_i^{\phantom{\dagger}} , \rho \right\}   - 2b_i^{\phantom{\dagger}} \rho b_i^\dagger  \right) \nonumber \\
    &\; - \sum_{n_z} \frac{\kappa_{n_z}}{2} \left(  \left\{ c_{n_z}^\dagger c_{n_z}^{\phantom{\dagger}} , \rho \right\}  - 2c_{n_z}^{\phantom{\dagger}} \rho c_{n_z}^\dagger    \right),
    \label{eq:Lindblad master equation}
\end{align}
where we consider two different phenomenological Markovian baths in order to take into account both the damping rates $\gamma_{i}$ of the dipole excitations, typically coming from radiative and Ohmic losses, and the damping rates $\kappa_{n_z}$ of the photons, arising from the imperfect cavity mirrors.
In the remainder of our paper, we assume that the dipolar losses are independent of the dipole site, such that $\gamma_i = \gamma$, and that the cavity losses are independent of the photon mode, i.e., $\kappa_{n_z} = \kappa$.
To study the transport properties along the chain of dipoles, we introduce the dimensionless dipole moment $p_i = \langle b_i^{\phantom{\dagger}} + b_i^\dagger \rangle$ bared by the dipole $i$.
As detailed in Appendix~\ref{sec:Stationary transport details}, we obtain from the master equation \eqref{eq:Lindblad master equation} the steady-state solution for the amplitudes $|p_i|$.

In the following, we fix the cavity loss to $\kappa/\omega_0 = 0.001$. 
This is motivated by the fact that all the qualitative propagation features found here are independent of the value of $\kappa$, as long as it remains small enough, as discussed in Appendix~\ref{sec:Cavity losses}.
The value of the dipolar damping rate $\gamma$, however, is of crucial importance and in order to illustrate its impact, we consider two different cases.
The first one, $\gamma/\omega_0 = 0.001$, can be achieved experimentally in platforms with small or highly controllable losses such as, e.g., microwave resonators or dielectric and SiC nanoparticles.
The second one, $\gamma/\omega_0 = 0.02$ represents the case of more lossy dipoles, which is naturally achieved in, e.g., nanoplasmonic setups where Ohmic losses are significant.

\subsection{Cavity-enhanced transport in an ordered chain}

In order to understand the underlying transport mechanisms of the system, we begin our study by considering the ordered case ($W/\omega_0=0$), which, due to the presence of polaritonic excitations, already features interesting transport properties.
Indeed, cavity photon excitations having an intrinsically collective and delocalized character, they are naturally propagating at longer distances. 
Moreover, the light-matter coupling Hamiltonian $\eqref{eq:H_dpph}$ acts as an effective long-range coupling between the dipoles.
Being in the Coulomb gauge, it can be seen as containing the retardation effects of the dipole-dipole interaction.
In the limit of an infinite cavity, i.e., in vacuum, it thus amounts to consider the usual $1/r$ long-range dipole-dipole coupling term. 
Such $1/r$ transport has been studied in the past in ordered \cite{Fung2011,BrandstetterKunc2016,Downing2017_Retardation} and disordered \cite{Markel} systems, showing the above-mentioned effect of long-range transport enhancement by the light-matter coupling.
In a finite cavity allowing for the strong-coupling regime, it has been shown in the case of a single mode Tavis-Cummings model \cite{Botzung2020,Chavez21,Dubail22} that under some approximations, the light-matter coupling can be expressed as a distance-independent hopping term.

\begin{figure}
    \centering
    \includegraphics[width=\columnwidth]{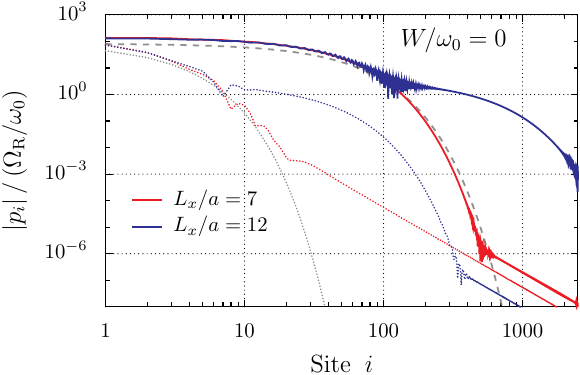}
    \caption{Steady-state amplitude of the dipole moment $|p_i|$ on site $i$ in units of the reduced Rabi frequency $\Omega_\mathrm{R}/\omega_0$ in an ordered ($W/\omega_0=0$) chain of $\mathcal{N}=2500$ dipoles, resulting from a monochromatic drive on the first dipole at a frequency $\omega_\mathrm{d}/\omega_0 = 1.0$, corresponding to a dark state.
    Results are shown for different cavity heights $L_x/a = 7$ (red lines, weak coupling) and $L_x/a = 12$ (blue lines, strong coupling), and different dipole losses $\gamma/\omega_0 = 0.001$ (solid lines) and $\gamma/\omega_0 = 0.02$ (dotted lines).
    The dashed (dotted) grey line represents the analytical estimate \eqref{eq:Propagation analytical} with $\gamma/\omega_0=0.001$ ($\gamma/\omega_0=0.02$).}    \label{fig:Propagation ordered different Lx}
\end{figure}

Here, to understand the effects of the strong-coupling regime, we begin our analysis by comparing transport properties in systems with different cavity heights.\footnote{Nevertheless, due to cavity confinement, the small heights $L_x/a \lesssim 9 $ that lead here to the weak-coupling regime are not comparable to the weak coupling of a dipole chain to vacuum electromagnetic modes, recovered here for an infinite cavity.}
To this end, in Fig.~\ref{fig:Propagation ordered different Lx} we show on a log-log scale the steady-state amplitude of the dipole moment $|p_i|$ along the sites $i$ of a chain of $\mathcal{N}=2500$ dipoles, for cavity heights $L_x/a=7$ (red lines) and $L_x/a=12$ (blue lines).
The amplitudes are given in units of $\Omega_\mathrm{R}/\omega_0$, since $|p_i| \propto \Omega_\mathrm{R}/\omega_0$.
Two different dipole losses are considered, $\gamma/\omega_0 = 0.001$ (solid lines) and $\gamma/\omega_0 = 0.02$ (dotted lines), and the result corresponds to the driving of a dark state at frequency $\omega_\mathrm{d}/\omega_0 = 1.0$.
In the figure, the dashed and dotted grey lines show the result in the case of a quasistatic dipole chain, without any light-matter coupling, and only with nearest-neighbor quasistatic dipole interaction.
Such decay has been computed analytically in Ref.~\cite{BrandstetterKunc2016} for the root-mean-square of the dipole moment.
Here, for the amplitude of the dipole moment it translates into
\begin{equation}
    |p_i| = \frac{2}{\pi}\frac{\Omega_\mathrm{R}}{\Omega_0} \left[ \sqrt{ 1 + \left(\frac{\gamma}{4\Omega_0}\right)^2  }  -  \frac{\gamma}{4\Omega_0}   \right]^i,
    \label{eq:Propagation analytical}
\end{equation}
such that it follows an exponential decay $|p_i| \propto \mathrm{e}^{-id_0/\zeta} $, with $i$ being the dipole site and $\zeta = d_0/\mathrm{arcsinh}\left( \gamma/4\Omega_0 \right)$ the propagation length.

Remarkably, we observe from Fig.~\ref{fig:Propagation ordered different Lx} that such an exponential decay is in very good agreement with the propagation of both the weak (red lines) and strong (blue lines) coupling regimes at short distances ($i \lesssim 100$) for small losses (solid lines), and at very short distances ($i \lesssim 10$) for larger losses (dotted lines).
Therefore, strong coupling does not modify the short-range propagation properties of the dark states, which are entirely attributable to the nearest-neighbor quasistatic dipole-dipole coupling.
The effect of the strong light-matter coupling becomes visible only at larger distances and takes the form of a second, less steep exponential decay, occurring after sharp oscillations that signal a change of regime.

After the exponential laws, the propagation follows an algebraic decay with slope $\sim 1/i^3$, visible as straight lines on the log-log scale of Fig.~\ref{fig:Propagation ordered different Lx}.
Such power law results from the $1/r^3$ quasistatic dipole-dipole coupling of the dipolar Hamiltonian \eqref{eq:H_dp}.
It is hence also visible in the quasistatic case, without light-matter interaction, when going beyond the nearest-neighbor approximation performed in Eq.~\eqref{eq:Propagation analytical}.
Moreover, this behavior has also been noticed in dipole chains coupled to vacuum electromagnetic modes \cite{Markel,Fung2011,BrandstetterKunc2016} and it also appears without quasistatic dipole-dipole but with nonzero light-matter coupling, the photons inducing an effective all-to-all dipolar coupling.
Importantly, such an algebraic decay leads the cavity-enhanced transport induced by strong coupling, that is, the second exponential regime, to be effective only for specific system sizes.
At too large distances, namely here with dipole losses $\gamma/\omega_0=0.02$ for $i \gtrsim 300$ (see the crossing between the blue and red dotted lines), the algebraic decay becomes dominant and the propagation is again similar for systems in the weak- and strong- coupling regimes.

\begin{figure}
    \centering
    \includegraphics[width=\columnwidth]{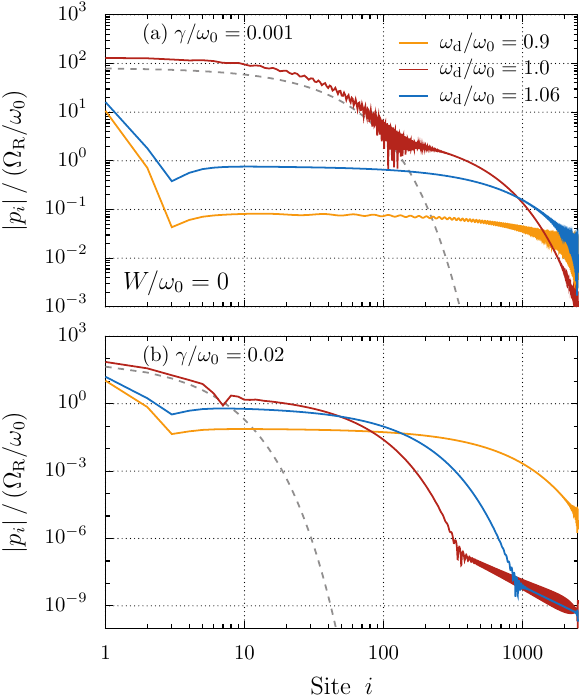}
   \caption{Same quantity as in Fig.~\ref{fig:Propagation ordered different Lx}, with dipole losses (a) $\gamma/\omega_0=0.001$ and (b) $\gamma/\omega_0=0.02$, but for a fixed cavity height $L_x/a=12$ and different driving frequencies $\omega_\mathrm{d}/\omega_0 = 0.9$, $1.0$, and $1.06$, exciting respectively a mainly photonic polariton, a dark state, and a mainly dipolar polariton. 
   The grey dashed lines show the analytical formula for $\omega_\mathrm{d}/\omega_0=1.0$ given in Eq.~\eqref{eq:Propagation analytical}.}
       \label{fig:Propagation ordered different frequencies}
\end{figure}

Now that we have studied the transport properties of a dark state in Fig.~\ref{fig:Propagation ordered different Lx}, we move to the propagation of polaritons.
To that purpose, in Fig.~\ref{fig:Propagation ordered different frequencies} we fix the cavity height to $L_x/a=12$ and we compare the propagation along the chain between driving frequencies $\omega_\mathrm{d}/\omega_0=0.9$, $1.0$, and $1.06$, which correspond, respectively, to a mainly photonic polariton, a dark state, and a mainly dipolar polariton.
Note that states with the same eigenfrequencies were already studied in the previous section, see Figs.~\ref{fig:Freq, PR and Ph function of W} and \ref{fig:Probability density for different disorder}, where the same color code was used.

By looking at Fig.~\ref{fig:Propagation ordered different frequencies}(a), where small dipole losses $\gamma/\omega_0 = 0.001$ are considered, it is clear that the amplitude of the dipole moment on the first site, $|p_1|$, is much smaller for polaritons (orange and blue lines) than for the dark state (red line).
This can be understood from the fact that the drive \eqref{eq:H_drive}, as well as the amplitude that we evaluate, are dipole-related quantities.
The dark eigenmodes can thus be more easily excited by the drive than the polaritons, and the maximum amplitude resulting from a drive at a given frequency is then related to the dipolar part of the associated eigenstate.
The first exponential is very steep for the polaritons, especially for the one which is mainly photonic, in orange, leading to an inefficient short-range transport. 
Indeed, the first exponential regime arising solely from the quasistatic dipole-dipole interaction, it is less efficient for a state whose photonic part is important, and for the polaritons it occurs on the first three sites only.
It is quickly supplanted by the second, photon-induced exponential which is less and less steep when the photonic part of the driven state increases (from red to blue to orange lines).
Such second exponential decay allows for the cavity-enhanced transport, and the long-range propagation of the polaritons becomes better than the one of the dark state from $i\gtrsim1000$.
We note that the slight increase between the two exponentials arises from the change of regime.

By increasing the dipole losses to $\gamma/\omega_0=0.02$ in Fig.~\ref{fig:Propagation ordered different frequencies}(b), we observe that, crucially, it is mainly the first, dipole-induced exponential and thus the dark state propagation that suffers from larger losses. 
Thereby, it leads to a strong dominance of the polaritons at long distances, thanks to the second, photon-induced exponential regime.
The mainly photonic polariton driven at $\omega_\mathrm{d}/\omega_0=0.9$, in orange, here leads to a dipole moment amplitude at the end of the chain, $|p_\mathcal{N}|$, which is $10^4$ times larger than the one of the dark state in red. Due to the algebraic decay, however, the propagation of the dark state catches up the ones of the polaritons at a long enough distance, leading, importantly, to a size-dependency of the cavity-enhanced transport.
For example, the mainly dipolar polariton (blue curve) shows a cavity-enhanced transport only from sites $i\sim50$ to $i\sim1000$.

To conclude this subsection, we have seen that the short-range transport is dominated by the nearest-neighbor quasistatic dipole-dipole coupling, that leads to a first exponential decay which provides efficient dark state propagation, but very poor polaritonic one.
The medium and long-range transport are, on the other hand, highly influenced by the strong light-matter coupling, which leads to the appearance of a second exponential decay regime showing the effect of cavity-enhanced transport.
The steepness of this photon-induced decay is flattened when the photonic part of the driven eigenstate increases, leading polaritons to propagate efficiently at long distances.
At larger distances, however, an algebraic decay independent of the nature of the eigenstate cancels this transport enhancement. Crucially, the increase of dipole losses lowers essentially the short-range propagation, and has less impact on the photon-induced second exponential decay regime.
This allows polaritons to better dominate medium and long-range transport when highly lossy dipoles are considered.

In the next subsection, we study the propagation along the chain in the presence of disorder.
We show that the same transport mechanisms are present, with however, the crucial addition of the disorder-induced mixing between dipolar and photonic weights, which we unveiled in Sec.~\ref{sec:Spectrum}.
Indeed, the dark states, that have been turned into semilocalized polaritonic states by disorder, will inherit the polaritonic propagation properties, that is, cavity-enhanced transport, that we discussed in this subsection.

\subsection{Disorder-enhanced transport}

\begin{figure}
    \centering
    \includegraphics[width=\columnwidth]{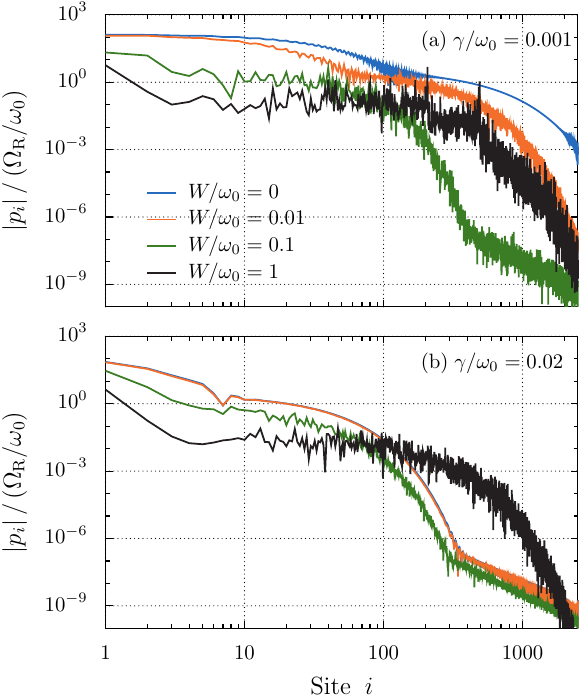}
    \caption{Same quantity as in Fig.~\ref{fig:Propagation ordered different Lx}, with driving frequency $\omega_\mathrm{d}/\omega_0=1.0$ corresponding to a dark state and dipole losses (a) $\gamma/\omega_0=0.001$ and (b) $\gamma/\omega_0=0.02$, but for a fixed cavity height $L_x/a=12$ and increasing disorder strength $W$. The data have been averaged over $100$ disorder realizations.}
    \label{fig:Propagation different disorder}
\end{figure}

The propagation along a disordered chain is presented in Fig.~\ref{fig:Propagation different disorder}, where the cavity height $L_x/a=12$.
We show on a log-log scale the steady-state amplitude of the dipole moment $|p_i|$ along the sites $i$ of the chain in units of the reduced Rabi frequency, resulting from a drive at frequency $\omega_\mathrm{d}/\omega_0=1.0$, for increasing disorder strength $W$.
In Fig.~\ref{fig:Propagation different disorder}(a), small dipole losses $\gamma/\omega_0=0.001$ are considered.
In such a case, we observe that increasing the disorder strength from $W/\omega_0=0$ (blue line) to $W/\omega_0=0.01$ (orange line) and $W/\omega_0=0.1$ (green line) suppresses more and more the transport, especially affecting the first exponential decay and the amplitude on the first site.
Increasing further the disorder to $W/\omega_0=1$ (black line), we observe an even weaker short-range transport, but a stronger long-range one.
Crucially, this long-range propagation arises from a flatter second exponential decay.
The steepness of the latter decay being related to the photonic weight of the driven eigenstate, this corroborates the disorder-induced gain of photonic weight of the dark states around $\omega_0$ that we unveiled in Fig.~\ref{fig:Freq, PR and Ph function of W}(c).
This demonstrates the link between the observed disorder-enhanced transport and the disorder-induced mixing between dark and photonic states, such that at strong  disorder strength (black line), the state driven at $\omega_\mathrm{d}/\omega_0=1.0$ is not anymore an almost uncoupled dark state but an hybridized polariton.
It thus propagates very similarly to polaritons in the absence of disorder (cf. the orange and blue lines in Fig.~\ref{fig:Propagation ordered different frequencies}).
We note that it also supports the observations made when looking at the localization of the eigenstates for increasing value of $W$ in Fig.~\ref{fig:Probability density for different disorder}, where panels (d), (e), and (f) correspond, respectively, to the orange, green, and black curves of Fig.~\ref{fig:Propagation different disorder}.

When increasing the dipoles losses to $\gamma/\omega_0=0.02$ in Fig.~\ref{fig:Propagation different disorder}(b), the dark state transport is strongly reduced while the polaritonic one remains almost unaffected. 
This enables an increasingly effective disorder-enhanced transport when the considered dipoles are lossy.
Here, it leads to a long-range propagation that is stronger with disorder than without, up to four orders of magnitudes around the sites $i\sim300$-$400$.

\begin{figure}
    \includegraphics[width=\columnwidth]{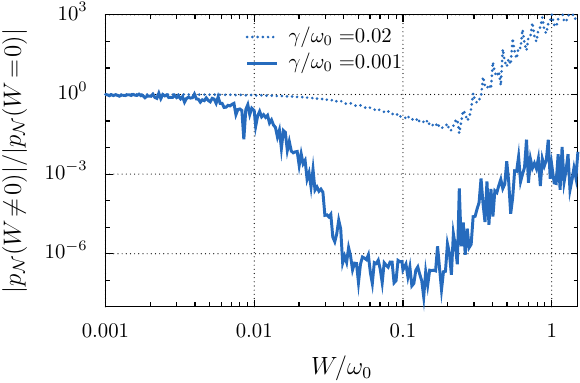}
    \caption{Steady-state amplitude of the last dipole moment of the chain, $|p_\mathcal{N}|$, normalized by the one obtained without disorder, and as a function of the disorder strength $W$ in units of $\omega_0$.
    A chain of $\mathcal{N}=500$ dipoles is considered, the driving frequency $\omega_\mathrm{d}/\omega_0=1.0$, the cavity height $L_x/a=12$, and the data have been averaged over $100$ disorder realizations.}
    \label{fig:Transport on last site versus disorder}
\end{figure}

To highlight the effect of disorder on long-range transport, we show in Fig.~\ref{fig:Transport on last site versus disorder} the evolution of the steady-state amplitude of the last dipole, $|p_\mathcal{N}|$, of a chain of $\mathcal{N}=500$ dipoles when the frequency disorder is increased.
We normalize the value of the amplitude $|p_\mathcal{N}|$ by the one at the end of an ordered chain, $|p_\mathcal{N}(W\!=\!0)|$.
Small dipole losses $\gamma/\omega_0=0.001$ (solid line) as well as larger ones $\gamma/\omega_0=0.02$ (dotted line) are considered, and the driving frequency is $\omega_\mathrm{d}/\omega_0=1.0$, just as in Fig.~\ref{fig:Propagation different disorder}.
While the long-range transport is only slightly affected at small disorder strength, it becomes suppressed when $W$ increases, in agreement with the eigenspectrum analysis of Sec.~\ref{sec:Spectrum}, where exponentially localized states were observed (see Fig.~\ref{fig:Probability density for different disorder}).
Then, increasing further the disorder strength leads to the rise of the amplitude on the last dipole site, that is, to the disorder-enhanced transport regime.
With small dipole losses (solid line), this increase does not exceed the value without disorder, remaining about $10^3$ times smaller.
But crucially, by considering more lossy dipoles (dotted line), an amplitude of the last dipole moment about $10^3$ times larger than without disorder can be observed with a disorder strength $W/\omega_0=1.5$.
We note that from $W/\omega_0=1$, the curves become essentially flat, indicating a regime where the amplitude is no longer affected by disorder.
This is in line with what has been termed disorder-independent transport in Ref.~\cite{Chavez21}, where a study comparable to ours was carried out for a disordered Tavis-Cummings model in a two-terminal, lossless transmission setup.
In our model, we explain this regime from the fact that once a state gained enough photonic weight to inherit an almost flat, efficient polaritonic transport up to the end of the chain, increasing further the disorder has no effect.

However, as well as the cavity-enhanced transport unveiled in Fig.~\ref{fig:Propagation ordered different frequencies}, in our model, the size of the chain $\mathcal{N}$ is critical to observe the disorder-enhanced and disorder-independent transport regimes. 
This is due to the complexity of the propagation which is made of different regimes of decay, and especially due to the algebraic $1/i^3$ tail appearing at very large distances.
For example, in Fig.~\ref{fig:Propagation different disorder}(b), the highly disordered case with $W/\omega_0=1$ (black line) shows an enhanced propagation only between the sites $i\simeq100$ and $i\simeq2000$.
To improve the transport at longer distances, a larger disorder strength should be considered, such that the photonic weight of the driven eigenstate further increases, leading to a flatter second exponential decay which can still dominate the algebraic decay at the end of the chain.
Remarkably, however, since the disorder parameter $W/\omega_0$ cannot be arbitrarily increased [see the discussion after Eq.~\eqref{eq:Dipolar coupling strength}], this points to the fact that the disorder-enhancement cannot be reached for too long chains by tuning only the frequency disorder.

\begin{figure}
    \includegraphics[width=\columnwidth]{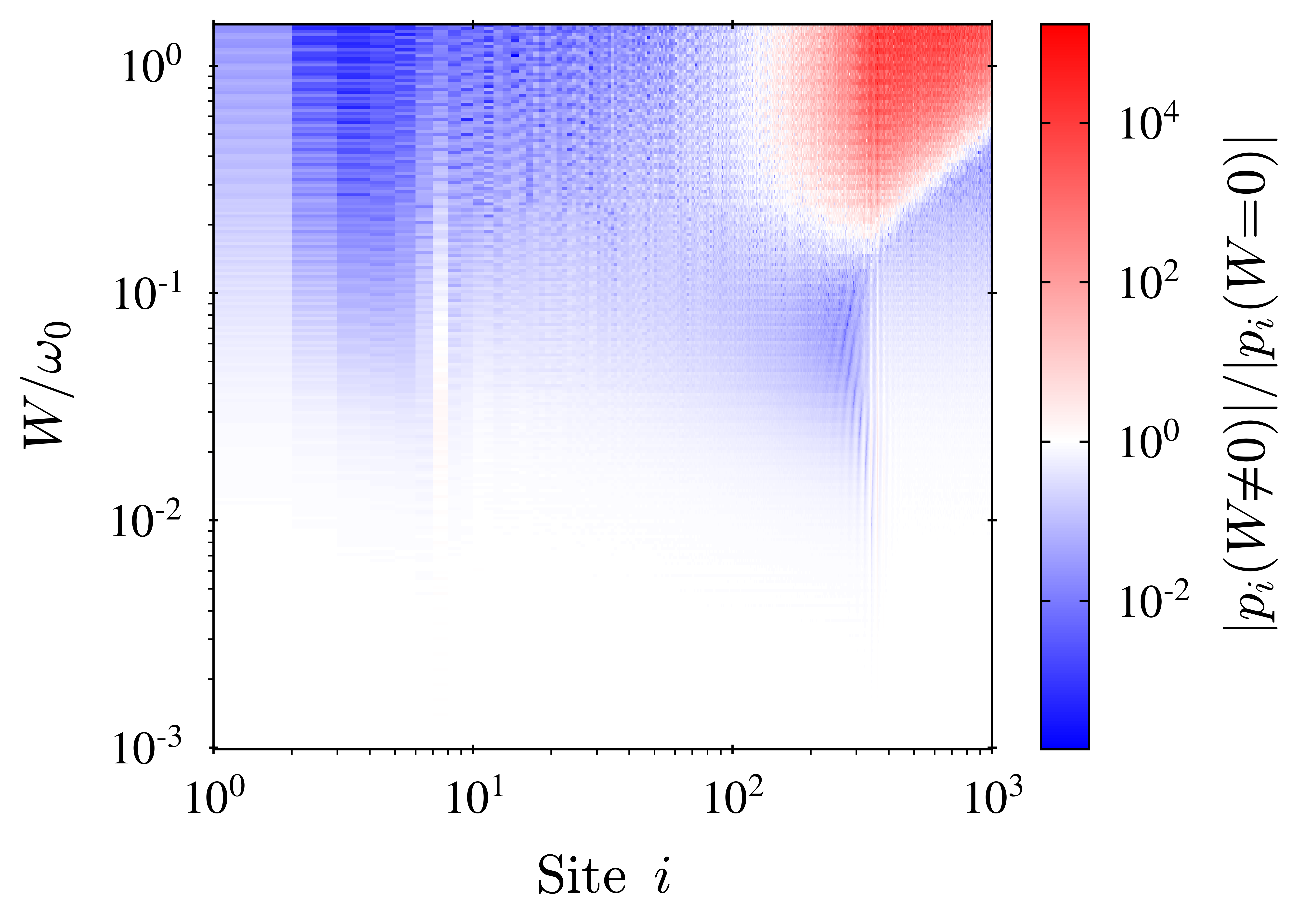}
    \caption{Steady-state amplitude of the dipole moment $|p_i|$, normalized by the one obtained without disorder, as a function of both the site $i$ along the chain and the disorder strength $W$ in units of $\omega_0$.
    The dipole losses $\gamma/\omega_0=0.02$, the driving frequency $\omega_\mathrm{d}/\omega_0=1.0$, the cavity height $L_x/a=12$, and the data have been averaged over 100 disorder realizations.}
    \label{fig:Density plot transport versus site versus disorder}
\end{figure}

We illustrate this effect of size-dependence in Fig.~\ref{fig:Density plot transport versus site versus disorder}, by computing the steady-state amplitude of the dipole moment normalized by the one without disorder, as a function of both the site $i$ along the chain and the disorder strength $W$.
The dipole losses $\gamma/\omega_0=0.02$ and the driving frequency $\omega_\mathrm{d}/\omega_0=1.0$, as in Fig.~\ref{fig:Propagation different disorder}(b) and in Fig.~\ref{fig:Transport on last site versus disorder} (dotted line).
In such density plot, one can distinguish between the different regimes of transport induced by disorder, according to the size of the chain, i.e., to the site $i$.
The white phase at the bottom of the plot shows that small disorder has no effect even on large chains, while the blue phase shows the Anderson localization regime, with a reduced transport.
The red phase, at the top right of the plot, indicates the disorder-enhanced transport, especially where the amplitude of the dipole moment is larger with disorder than without.
At short distances, such red phase is absent, showing that no disorder-enhanced transport is achievable for small chains with the parameters considered here.
For larger distances, one remarks that the disorder required to enter in the red phase goes linearly with the distance $i$, implying that a larger chain needs a larger disorder strength.

Furthermore, it is interesting to note that decreasing the dipolar coupling $\Omega_0$ can enhance the disorder-induced effects (not shown), allowing for a disorder-enhanced transport for smaller disorder strength $W$, and operating at shorter distances.
In view of our model, we understand this by the fact that it would lead to a suppression of the first exponential decay, but would have less impact on the second one, the latter being mediated by cavity photons.
The relative dominance of the polaritonic long-range transport over the dark state one would thus be larger, in a similar fashion as with the increase of dipole losses.
At specific distances, a very small dipolar coupling can also lead to a revival of Anderson localization appearing after the disorder-enhanced and disorder-independent regimes, similarly to what was observed in Ref.~\cite{Chavez21}.
However, decreasing the dipolar coupling, i.e., increasing the spacing $d/a$ between the dipoles, would also go along with a drastic overall reduction of the propagation \cite{BrandstetterKunc2016}.

\begin{figure*}[t]
    \centering
    \includegraphics[width=\linewidth]{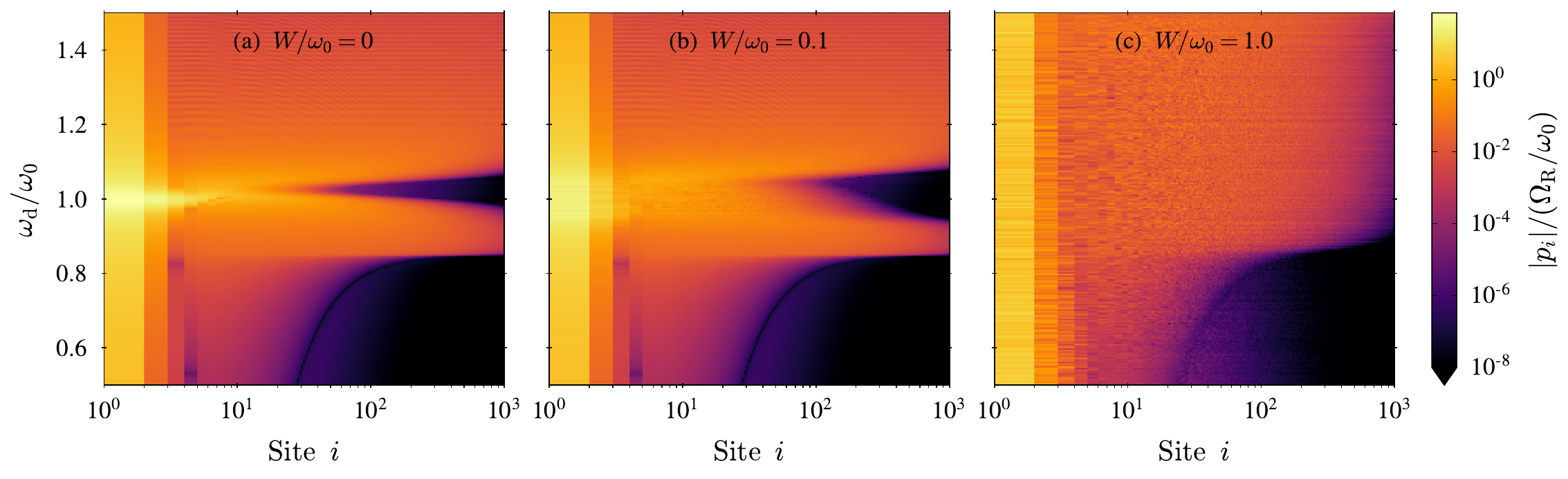}
    \caption{Density plots of the steady-state amplitude of the dipole moment $|p_i|$ in units of the reduced Rabi frequency $\Omega_\mathrm{R}/\omega_0$, as a function of both the site $i$ and the driving frequency $\omega_\mathrm{d}/\omega_0$.
    Three cases are presented: (a) Without disorder, (b) $W/\omega_0 = 0.1$, and (c) $W/\omega_0 = 1$.
    The dipole losses $\gamma/\omega_0=0.02$, the cavity height $L_x/a=12$, and the data have been averaged over $100$ disorder realizations.}
    \label{fig:Heatmap transport}
\end{figure*}

Finally, to study the effect of driving frequencies $\omega_\mathrm{d}/\omega_0\neq1.0$, we show on Fig.~\ref{fig:Heatmap transport} the steady-state amplitudes of the dipole moments in a density plot, as a function of both the driving frequency and the site $i$ along a chain of $\mathcal{N}=1000$ dipoles.
On panels (a), (b), and (c), we compare the propagation in a chain with disorder strengths $W/\omega_0=0$, $W/\omega_0=0.1$, and $W/\omega_0=1$, respectively.
We consider lossy dipoles, with $\gamma/\omega_0=0.02$, for the purpose of maximizing the dominance of the long-range polaritonic transport over the dark state one.
In Fig.~\ref{fig:Heatmap transport}(a), without disorder, a bright yellow spot around $\omega_\mathrm{d}/\omega_0=1.0$ demonstrates that the maximal amplitude is generated at short range by driving a dark state, supporting the fact that these states are more easily driven than polaritons.
They present however inefficient long-range propagation (see the purple and black spots).
At the bottom of the spectrum, a clear threshold can be seen around $\omega_\mathrm{d}/\omega_0=0.85$, corresponding to the minimum of the LP branch.
The driven states below this threshold are not anymore eigenmodes of the system and hence present poor transport properties.
Finally, the orange strips above and below the bright spot around $\omega_\mathrm{d}/\omega_0=1.0$ correspond to hybridized polaritons, with poor short-range propagation but very good long-range one, as already observed in Fig.~\ref{fig:Propagation ordered different frequencies}.

By adding disorder, in Fig.~\ref{fig:Heatmap transport}(b), we clearly see the expansion of the effective dark state band around $\omega_\mathrm{d}/\omega_0=1.0$.
This expansion goes along with a decrease of the transport efficiency, with the disappearance of the bright yellow spot at short range and the enlargement of the black spot at longer range.
Polaritons outside the dark state band (orange strips), on the other hand, remain not affected by disorder, displaying the effect of cavity protection of their transport against disorder.
When further increasing the disorder strength, in Fig.~\ref{fig:Heatmap transport}(c), all the frequencies up to $\omega_\mathrm{d}/\omega_0=1.5$ belong to the dark state band. 
All these now hybridized and semilocalized states have approximately the same propagation characteristics.
This, crucially, leads to an improved long-range transport of the dark states around $\omega_\mathrm{d}/\omega_0=1.0$, which can be readily seen from the complete disappearance of the black spot.
We note that it also leads to a slightly enhanced short-range transport of the mainly photonic polaritons, the latter gaining a dipolar weight from the disorder-induced hybridization, allowing them to be more easily driven. 
Their long-range transport is however reduced by disorder.
In agreement with what was observed in Sec.~\ref{sec:Spectrum} through the fully dipolar and exponentially localized states visible as the black regions in Figs.~\ref{fig:Freq, PR and Ph function of W}(a) and \ref{fig:Freq, PR and Ph function of W}(b), the eigenstates driven with a frequency essentially smaller than the lowest bare photonic state $\omega_{q=0}^\mathrm{ph}/\omega_0 \simeq 0.85$ do not inherit polaritonic propagation features, the light-matter coupling being not strong enough to let them be coupled to photons by the disorder.

\section{Conclusions}
\label{sec:Conclusion}

We provided a study of the interplay between light-matter coupling, dipolar coupling, and disorder, through the analysis of both the localization properties and the transport along a chain of disordered dipoles placed inside an optical cavity.
We considered a multimode light-matter coupling Hamiltonian which goes beyond the typical single-mode coupling widely used in quantum optics, as well as an all-to-all $1/r^3$ quasistatic dipole-dipole coupling.
Crucially, losses in both the dipoles and the mirrors are taken into account.
The disorder was considered on the individual resonance frequencies $\omega_i$ of the dipoles.

In our multimode model, all the eigenstates are hybridized by the strong light-matter coupling.
However, we can distinguish the \emph{dark states}, with eigenfrequencies around the average bare dipole resonance frequency $\omega_0$ and almost zero photonic weight, and the \emph{polaritons}, which have an eigenfrequency highly renormalized by the coupling and a significant photonic weight.
We showed that at weak and medium disorder strengths, dark states suffer from Anderson localization, becoming exponentially localized on a few sites of the chain.
On the other hand, the polaritons are more robust against frequency disorder, and show a cavity-protection effect \cite{Houdre1996,Michetti_2005}.
Notably, the only states impacted by disorder are the ones located in the same frequency window as the possible disordered individual dipole frequencies $\omega_i$, namely around $\omega_0 \pm W/2$.
Increasing the disorder strength leads this frequency window to act as an effective dark state band, whose bandwidth increases at a rate $W/2$.

At large disorder strength, we have shown that the states located both in this dark state band and with an eigenfrequency larger than the lowest bare photonic mode $\omega_{q=0}^\mathrm{ph}$ are subject to a disorder-induced hybridization with photons, allowing new coupling between matter and photonic degrees of freedom.
Indeed, the photonic part of these states increases linearly with the disorder strength and such states become semilocalized \cite{Botzung2020}, i.e., localized in multiple nonadjacent sites.
The eigenfrequencies smaller than $\omega_{q=0}^\mathrm{ph}$, on the other hand, cannot be hybridized through disorder, and suffer from the usual Anderson exponential localization.

By studying the transport of an excitation along the chain in a driven-dissipative scenario, we have shown that the strong light-matter coupling regime leads to a second exponential decay in the propagation, following a first one which is solely due to nearest-neighbor quasistatic dipole-dipole interactions.
The larger the photonic part of the driven state is, the poorer the short-range propagation is, but the more flat this second exponential regime is.
At large frequency disorder, the dark states, which have become polaritons through the above-mentioned mechanism of disorder-induced hybridization, inherit polaritonic transport properties, namely a poor short-range transport but an efficient long-range one.
Moreover, we have found that this disorder-enhanced transport is not efficient over too long distances, or at least requires very large disorder strength. Indeed, the amount of disorder required to enter into such a regime increases linearly with the system size.

Crucially, we have shown that increasing the dipole losses impacts essentially the short-range transport, mediated by the quasistatic dipole-dipole coupling, which consequently increases the dominance of the polaritons at longer range.
In this way, we have found that by considering highly lossy dipoles, a large frequency disorder can increase the long-range transport up to a factor $10^4$ as compared to the case without disorder, see, e.g., around the site $i=400$ in Fig.~\ref{fig:Propagation different disorder}(b).
In terms of the power radiated by the dipoles, it translates into an increase of up to $8$ orders of magnitude [see Eq.~\eqref{eq:Larmor}].
This drastic increase of long-range transport could be useful especially for setups using plasmonic nanoparticle chains, where experiments have witnessed short transport distances \cite{Krenn1999,Maier2002,Maier2003,Koenderink2007,Crozier07,Apuzzo2013,Barrow2014}, notably due to large Ohmic losses.
Indeed, when considering plasmonic nanoparticles, the dimensionless dipole strength $k_0a=0.1$ used throughout this work corresponds to nanoparticles with, e.g., radius $a=\SI{7.6}{\nm}$ and average resonance frequencies $\omega_0=\SI{2.6}{\eV\per\hbar}$. Considering a driving field intensity $I_0 = \SI{1}{\MW\per\cm^2}$, corresponding to, e.g., a laser of power $\mathcal{P}_0=\SI{7.1}{\mW}$ focused on a spot of size $\pi\lambda_0^2 \simeq \SI{0.71}{\mu\meter^2}$,
the power radiated in the far field by a plasmonic nanoparticle with decay rate $\gamma/\omega_0=0.02$ around the site $i=400$ [see Fig.~\ref{fig:Propagation different disorder}(b)] can then be increased from $\mathcal{P}_{400}(W/\omega_0=0) \simeq \SI{e-24}{\W}$ to $\mathcal{P}_{400}(W/\omega_0=1) \simeq \SI{e-16}{\W}$.
While the ordered case leads to a very small power which is clearly experimentally unreachable, a setup with sub-femtowatt sensitivity should detect the radiated power at the end of a highly disordered plasmonic chain.

Our model, which allowed to study the particularly nontrivial interplay between disorder and light-matter coupling, opens the way to study more complex systems which could be of larger dimensionality, or feature topological properties \cite{Mann2018,Downing2019,Mann2020,Mann2022}, for which the interplay with disorder and a coupling beyond nearest neighbor has recently presented interesting phenomena \cite{Perez-Gonzalez2019}.
Moreover, there is a growing interest in topological photonics \cite{Ozawa2019}, which one of the key challenges is to combine disorder-robust topological edge states with strong light-matter coupling.
A detailed understanding of the interplay between disorder and strong
light-matter coupling, allowed by our model, would then be of crucial importance to this field.

\textit{Note added.} Recently, Ref.~\cite{Cao_arXiv2022}, which considers a similar system as ours, in particular a multimode photonic cavity, appeared on the arXiv.


\begin{acknowledgments}
We  would  like  to  acknowledge Stéphane Berciaud, Charles A.\ Downing, David Hagenm\"uller, Rodolfo A. Jalabert, Gaëtan Percebois, and Dietmar Weinmann for insightful discussions.
This work of the Interdisciplinary Thematic Institute QMat, as part of the ITI 2021-2028 program of the University of Strasbourg, CNRS, and Inserm, was supported by IdEx Unistra (ANR 10 IDEX 0002), and by SFRI STRAT’US Projects No.\ ANR-20-SFRI-0012 and No.\ ANR-17-EURE-0024 under the framework of the French Investments for the Future Program.
\end{acknowledgments}

\appendix

\section{Analytical diagonalization in the case of an infinite ordered chain}
\label{sec:Diagonalization}

In this Appendix, we reproduce for the sake of self-containedness the diagonalization procedure of Ref.~\cite{Downing2021} in the case of an infinite ordered dipole chain within a cuboidal photonic cavity, and give analytical expressions for the eigenfrequencies and eigenstates of the system.
As in the main text, we here employ the rotating wave approximation.
In such translationally invariant case, one can use periodic boundary conditions, and it is then convenient to use the Fourier transform $b_j = \mathcal{N}^{-1/2}\sum_q\mathrm{e}^{\mathrm{i}jqd_0}b_q$ ($j=1,\dots,\mathcal{N}$), where the dipolar wavenumber $q = 2\pi p/\mathcal{N}d_0$, with the integer $p \in [-\mathcal{N}/2, +\mathcal{N}/2]$.
Importantly, the periodicity of the system in the $z$ direction implies that the photonic wavenumber $k_z=2\pi m_z/L_z$, $m_z\in\mathbb{Z}$, where $L_z=(\mathcal{N}+1)d_0$, is conserved with the dipolar one. In the following, we then write $k_z=q$.

By quantizing the vector potential \eqref{eq:Vector potential} in a cavity with periodic boundary conditions in the $z$ direction \cite{Kakazu1994}, and by rewriting the dipolar Hamiltonian \eqref{eq:H_dp} in Fourier space, the polaritonic Hamiltonian \eqref{eq:Hamiltonian approximated} becomes
\begin{equation}
    \tilde{H}^\mathrm{pbc} = \sum_q \mathbf{\phi}_q^\dagger \mathcal{H}_q \mathbf{\phi}_q, \;\;\;\;\;\;\;\;\; \mathcal{H}_q = \hbar
    \begin{pmatrix}
        \omega_q^\mathrm{dp} & \mathrm{i}\xi_q \\
        -\mathrm{i}\xi_q & \omega_q^\mathrm{ph}
    \end{pmatrix},
\label{eq:Hamiltonian polariton pbc}
\end{equation}
where $\mathcal{H}_q$ is the Bloch Hamiltonian, while $\mathbf{\phi}_q^\dagger = (b_q^\dagger,c_q^\dagger)$.
In the above expression, $\omega_q^\mathrm{dp}$ is the quasistatic dispersion of the collective dipolar excitation, i.e., the eigenspectrum of the dipolar Hamiltonian \eqref{eq:H_dp}.
It reads
\begin{equation}
    \omega_q^\mathrm{dp} = \omega_0 + \Omega_0f_q,
\label{eq: Dipole dispersion RWA Fourier}
\end{equation}
with the dipolar coupling strength $\Omega_{0} = (\omega_0/2)(a/d)^3$, and where the lattice sum $f_q$ can be expressed in closed form in terms of the polylogarithm function $\text{Li}_s(z) = \sum_{n=1}^{\infty}z^n/n^s$ as $ f_q^{\sigma} = \mathrm{Li}_3(\mathrm{e}^{\mathrm{i}qd}) + \mathrm{Li}_3(\mathrm{e}^{-\mathrm{i}qd})$.
Within such periodic boundary conditions, the photon dispersion \eqref{eq:Photon dispersion approximated} is
\begin{equation}
    \omega^{\mathrm{ph}}_q = c\sqrt{ \left( \frac{\pi}{L_y} \right)^2 + q^2 },
\label{eq:Photon dispersion approximated Fourier}
\end{equation}
and the light-matter coupling \eqref{eq:Light-Matter coupling} now reads
\begin{equation}
    \xi_q = \omega_0\sqrt{ \frac{2\pi a^3}{L_xL_yd} \frac{\omega_0}{\omega_q^\mathrm{ph}} }.
\label{eq:Light-matter coupling fourier}
\end{equation}

Using a bosonic Bogoliubov transformation, one can readily diagonalize the Hamiltonian \eqref{eq:Hamiltonian polariton pbc} as
\begin{equation}
    \tilde{H}^\mathrm{pbc} = \sum_{q,\tau} \omega_{q\tau}^\mathrm{pol} \beta_{q\tau}^\dagger \beta_{q\tau},
\label{eq:Hamiltonian polariton diagonalized}
\end{equation}
where the upper ($\tau=+$) and lower ($\tau=-$) polaritonic bands are given by
\begin{equation}
    \omega_{q\tau}^\mathrm{pol} = \frac{1}{2}\left( \omega_q^{\mathrm{ph}} + \omega_q^{\mathrm{dp}} \right) + \tau\sqrt{ \xi_q^2 + \Delta_q^2 },
\label{eq:Dispersion polariton}
\end{equation}
with $\Delta_q = (\omega_q^{\mathrm{ph}} - \omega_q^{\mathrm{dp}})/2$ being the light-matter detuning between the bare photonic and dipolar dispersions.

Finally, the Bogoliubov operators diagonalizing the Hamiltonian \eqref{eq:Hamiltonian polariton pbc} are a linear combination of the dipolar and photonic ladder operators, $\beta_{q\tau} = u_{q\tau}b_q + v_{q\tau}c_q$.
The modulus squared of the two coefficients $u_{q\tau}$ and $v_{q\tau}$, which are normalized as $|u_{q\tau}|^2 + |v_{q\tau}|^2 = 1$, thus represent, respectively, the dipolar part $\mathrm{D}_{q\tau}$ and the photonic part $\mathrm{Ph}_{q\tau}$ of the polaritonic eigenmodes.
These latter quantities read
\begin{subequations}
\begin{equation}
    \mathrm{D}_{q\tau} = |u_{q\tau}|^2 = \frac{1}{2}\left( 1 - \tau \frac{\Delta_q}{\sqrt{\xi_q^2 + \Delta_q^2}} \right)
\end{equation}
and
\begin{equation}
    \mathrm{Ph}_{q\tau} = |v_{q\tau}|^2 = \frac{1}{2}\left( 1 + \tau \frac{\Delta_q}{\sqrt{\xi_q^2 + \Delta_q^2}} \right).
    \label{eq:Photonic part fourier}
\end{equation}
\label{eq:Dipolar and photonic part fourier}
\end{subequations}

\section{Beyond the rotating wave approximation}
\label{sec:RWA}

Here, we provide justifications for the rotating wave approximation (RWA) used in the main text.
To this end, we consider the effects of counter-rotating terms in both the eigenvalues and transport properties of an ordered chain.

\begin{figure}[tbh]
    \centering
    \includegraphics[width=\columnwidth]{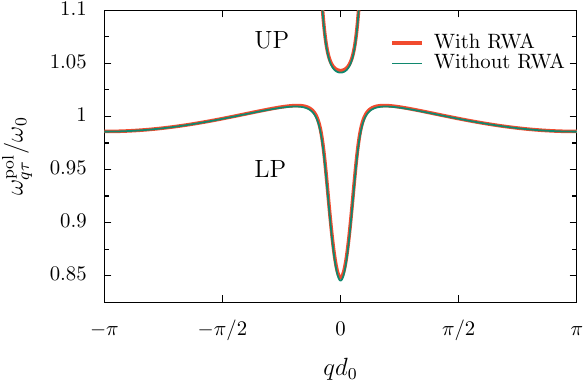}
    \caption{Polaritonic band structure of an ordered chain in the thermodynamic limit, in units of the bare frequency $\omega_0$, and as a function of the reduced wavenumber $qd_0$ in the first Brillouin zone. 
    In orange, we show the dispersion used in the main text, which is obtained using the RWA and given by Eq.~\eqref{eq:Dispersion polariton}, and we compare it to the one obtained without using the approximation, given by Eq.~\eqref{eq:Dispersion Fourier no RWA} and shown in green.
    The considered cavity height is $L_x/a=12$.
    }
    \label{fig:RWA dispersion}
\end{figure}

Considering the counter-rotating terms, the polaritonic Hamiltonian \eqref{eq:Hamiltonian approximated} that we use in the main text transforms into
\begin{align}
    \tilde{H}^\mathrm{CR} =&\, H_\mathrm{dp}^\mathrm{CR} + \sum_{n_z=1}^{\mathcal{N}} \hbar\omega^{\mathrm{ph}}_{n_z}{c^\dagger_{n_z}} c^{\phantom{\dagger}}_{n_z} \nonumber \\ 
    &\, + \mathrm{i}\hbar \sum_{i=1}^\mathcal{N} \sum_{n_z=1}^\mathcal{N} \xi_{in_z}\left( b_i^\dagger  - b_i^{\phantom{\dagger}} \right) \left( c_{n_z}^\dagger + c^{\phantom{\dagger}}_{n_z} \right),
\label{eq:Hamiltonian approximated no RWA}
\end{align}
where the dipolar Hamiltonian with counter-rotating terms reads \cite{Downing2017_Retardation}
\begin{equation}
    H^\mathrm{CR}_\mathrm{dp} = \sum_{i=1}^\mathcal{N} \hbar\omega_i b_{i}^{\dagger} b_{i}^{\phantom{\dagger}} + \sum_{i=1}^{\mathcal{N}-1} \sum_{j=i+1}^\mathcal{N} \hbar\Omega_{ij} \left( b_i^{\phantom{\dagger}} + b_i^\dagger \right) \left( b_{j}^{\phantom{\dagger}} + b_{j}^{\dagger} \right).
\label{eq:H_dp no RWA}
\end{equation}
In the limit of an ordered chain, similarly to what has been done in Appendix~\ref{sec:Diagonalization}, one can rewrite the Hamiltonian \eqref{eq:Hamiltonian approximated no RWA} in Fourier space using periodic boundary conditions and diagonalize it using a Hopfield-Bogoliubov transformation.
The diagonalization leads to the polaritonic dispersion
\begin{equation}
    w^\mathrm{pol}_{q\tau} = \sqrt{     \Gamma_q^2 + \tau \sqrt{ \Gamma_q^4 -( w_q^\mathrm{dp} )^2\left[  ( \omega_q^\mathrm{ph} )^2 - 4\frac{\xi_q^2}{\omega_0}\omega_q^\mathrm{ph}  \right] }     },
   \label{eq:Dispersion Fourier no RWA}
\end{equation}
where we have introduced the quantity
\begin{equation}
    \Gamma_q^2 = \frac{1}{2}\left[ \left( \omega_q^\mathrm{ph} \right)^2 + \left( w_q^\mathrm{dp} \right)^2 \right],
\end{equation}
with the bare photonic dispersion given in Eq.~\eqref{eq:Photon dispersion approximated Fourier}, and where 
\begin{equation}
    w_q^\mathrm{dp} = \omega_0\sqrt{ 1 + 2f_q\frac{\Omega_0}{\omega_0} }
\label{eq:Dipolar dispersion no RWA}
\end{equation}
is the dispersion of the dipolar Hamiltonian with counter-rotating terms \eqref{eq:H_dp no RWA}.
Details on this diagonalization procedure can be found in the supplementary material of Ref.~\cite{Downing2021}, with the only difference that here we neglect the diamagnetic $A^2$-term in the light-matter Hamiltonian \eqref{eq:H_dpph}.

In Fig.~\ref{fig:RWA dispersion}, we show a comparison between the polaritonic dispersions obtained with (orange line) and without (green line) the RWA, given respectively by Eqs.~\eqref{eq:Dispersion polariton} and \eqref{eq:Dispersion Fourier no RWA}.
We observe that the two band structures overlap almost perfectly, with very slight differences noticeable only around the center of the BZ, where the light-matter coupling is maximal.
This is in agreement with the fact that our model does not allow the ultra-strong coupling (USC) regime to be reached, since one enters into the latter regime when the effects of the counter-rotating terms  become sizable \cite{FornDiaz2019}, which is clearly not the case here.
The fact that we remain in the strong-coupling regime only also justifies neglecting the diamagnetic $A^2$-term in the light-matter coupling Hamiltonian \eqref{eq:H_dpph}.
As a matter of fact, perceptible differences in the eigenspectrum caused by the absence of such diamagnetic term  were found in the context of the ultra-strong or deep-strong coupling regime only \cite{Stefano2019,Kockum2019_review}.

\begin{figure}[t]
    \centering
    \includegraphics[width=\columnwidth]{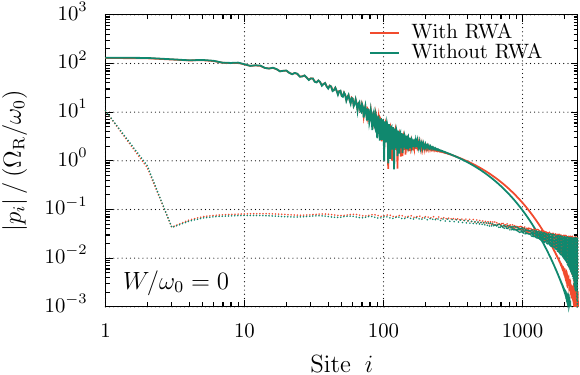}
    \caption{Steady-state amplitude of the dipole moment $|p_i|$ in units of the reduced Rabi frequency $\Omega_\mathrm{R}/\omega_0$ along the sites $i$ of an ordered chain.
    The results are shown within (orange) and without (green) the RWA.
    Two different driving frequencies are considered: $\omega_\mathrm{d}/\omega_0=1.0$, which drives a dark state is shown as solid lines, and $\omega_\mathrm{d}/\omega_0=0.9$, which drives a polariton with a large photonic part is shown as dotted lines. 
    The cavity height $L_x/a=12$, and the dipole and mirror losses are respectively $\gamma/\omega_0 = 0.001$ and $\kappa/\omega_0=0.001$.}
    \label{fig:RWA transport}
\end{figure}

\begin{figure*}[tbh]
    \includegraphics[width=\linewidth]{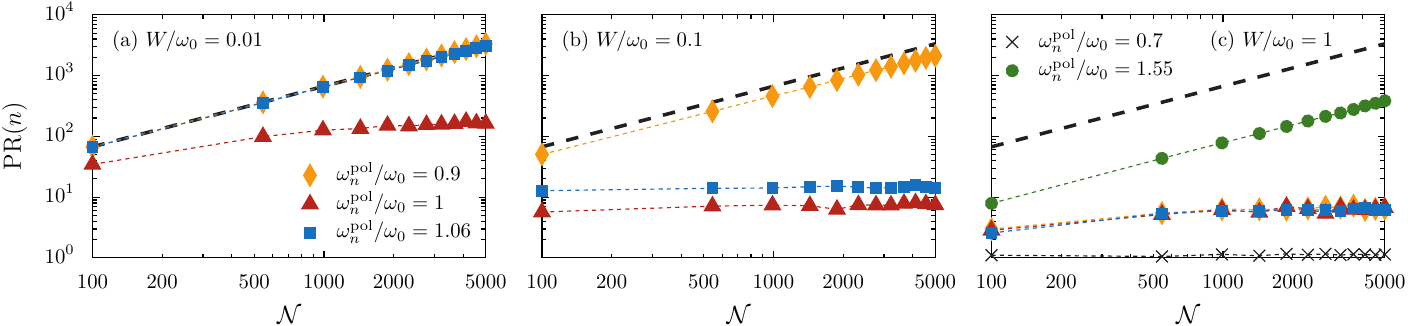}
    \caption{Participation ratio $\mathrm{PR}(n)$ [defined in Eq.~\eqref{eq:Participation Ratio}] as a function of the size of the chain $\mathcal{N}$, for the eigenmodes associated to the eigenfrequencies $\omega_n^\mathrm{pol}/\omega_0 = 0.9$, $1.0$, and $1.06$, with increasing disorder strength from the left to the right panel.
    Panel (c) shows in addition the result for the eigenmodes associated to the eigenfrequencies $\omega_n^\mathrm{pol}/\omega_0 = 0.7$ and $1.55$.
    Extended states show a participation ratio growing linearly with $\mathcal{N}$, while localized states present a constant participation ratio.
    The black dashed line represents the maximum growth rate $2(\mathcal{N}+1)/3$.
    The cavity height $L_x/a=12$, and the data have been averaged over $100$ disorder realizations.}
\label{fig:Finite size scaling}
\end{figure*}

However, in the case of driven-dissipative transport computations, special attention has to be taken when using the RWA.
Indeed, it has been shown \cite{Agarwal1971,Agarwal1973,West1984,Ford1997,Fleming2010} that using the RWA along with a Lindblad master equation can lead to inaccurate dynamics, especially when dropping the counter-rotating terms directly in the Hamiltonian, as we have done in the main text. 
In Fig.~\ref{fig:RWA transport}, we thus compare our transport results obtained with (orange lines) and without (green lines) neglecting the counter-rotating terms in the Hamiltonian entering in the Lindblad master equation \eqref{eq:Lindblad master equation}.
We show on a log-log scale the amplitude of the steady-state dipole moment $|p_i|$ in units of the reduced Rabi frequency $\Omega_\mathrm{R}/\omega_0$, as a function of the dipole site $i$ along the chain, when driving a dark state with $\omega_\mathrm{d}/\omega_0 = 1.0$ (solid lines), and a mainly photonic polariton with $\omega_\mathrm{d}/\omega_0=0.9$ (dotted lines).
As can be seen from Fig.~\ref{fig:RWA transport}, the propagation is qualitatively the same within and without the RWA.
The small differences arise essentially from the very slight frequency shift visible in Fig.~\ref{fig:RWA dispersion}, implying the fact that using a given driving frequency $\omega_\mathrm{d}$ does not drive an eigenstate with exactly the same photonic weight in the case of employing or not the RWA in the computation.


\section{On the nature of the semilocalized eigenstates}
\label{sec:multifractality}

In the main text, we show that at large disorder strength the eigenstates of the Hamiltonian \eqref{eq:Hamiltonian approximated} can be of three very different types [cf.\ Figs.~\ref{fig:Freq, PR and Ph function of W}(a) and \ref{fig:Freq, PR and Ph function of W}(b)]: (i) dipolar with a PR of order $1$, (ii) hybridized with a PR of order $10$, (iii) photonic with a PR of order $\mathcal{N}$.
By inspection of the probability density $|\Psi_i|^2$ of these states along the sites $i\in[1,\mathcal{N}]$ of the chain, we find that the dipolar ones are exponentially localized in the chain, with very small power-law tails, so that we coin them as \emph{localized}.
The photonic states are homogeneously extended along the chain, and we term them \emph{delocalized} at the scale of the system size, since their localization length is larger than the size of the chain.

The hybridized states, however, are localized on multiple, noncontiguous sites [see panels (c), (f), and (i) of Fig.~\ref{fig:Probability density for different disorder}, where $\mathcal{N}=1000$], so that we coin them  \emph{semilocalized}, following Ref.~\cite{Botzung2020} were similar features were observed in a disordered Tavis-Cummings (TC) model.
In the context of disordered systems, such neither localized nor delocalized states have been associated to multifractal, nonergodic extended states, which are notably a feature of the critical point of Anderson transitions \cite{Mirlin_RevModPhys}. Multifractal states have also been observed in disordered Floquet systems \cite{Roy2018} as well as in long-ranged disordered systems \cite{Levitov1999,Deng2016}.
Recently, Dubail \textit{et al.}\ \cite{Dubail22} unveiled the multifractal nature of the semilocalized states present in a disordered TC model without short-range hopping, where the light-matter coupling acts as an effective long-range hopping term.

In this Appendix, we study the nature of the semilocalized states present in our multimode light-matter coupling model, to ascertain whether they have the same properties as in simplified single-mode models, or if their particular localization profile comes from finite-size effects.
To this end, we first conduct a scaling analysis of the participation ratio with the system size $\mathcal{N}$, then we perform a level statistics analysis, before we carry out a multifractal analysis through the scaling of the generalized participation ratio with the system size.

\subsection{Scaling of the participation ratio}

We show in Fig.~\ref{fig:Finite size scaling} the scaling of the participation ratio $\mathrm{PR}(n)$ [defined in Eq.~\eqref{eq:Participation Ratio}] with the number of dipoles $\mathcal{N} \in [100,5000]$, for increasing values of the disorder strength $W/\omega_0=0.01$, $0.1$, and $1$ from the left to the right panel.
We note that to minimize fluctuations, all the results have been averaged over $100$ disorder realizations.
Such scaling is of crucial importance to correctly conclude on the localization nature of the eigenstates.
Indeed, the PR of a localized state must remain independent of the system size, while the one of an extended state should scale with the system size.

In Figs.~\ref{fig:Finite size scaling}(a) and \ref{fig:Finite size scaling}(b), we display the results for the eigenstates corresponding to the eigenfrequencies $\omega_n^\mathrm{pol}/\omega_0 = 0.9$, $1$, and $1.06$, i.e., the same as the ones studied in the main text, notably in Figs.~\ref{fig:Freq, PR and Ph function of W} and \ref{fig:Probability density for different disorder}.
On the one hand, at small disorder strength in panel (a),  two polaritonic states (plotted by orange diamonds and blue squares) show a PR increasing linearly with the number of dipoles $\mathcal{N}$, precisely following the growth rate $2(\mathcal{N}+1)/3$ plotted as a black dashed line, which corresponds to maximally extended eigenstates in an ordered ($W=0$) system.
On the other hand, the PR of the dark state, plotted by red triangles, increases at small system sizes before it converges to a constant value for large enough number of dipoles $\mathcal{N}$, demonstrating its localized nature despite its quite large localization length.
At larger disorder strength in panel (b), the PR of both the mainly dipolar polariton (blue squares) and the dark state (red triangles) are drastically reduced and remain constant, indicating localized states, as already seen in Figs.~\ref{fig:Probability density for different disorder}(e) and \ref{fig:Probability density for different disorder}(h).
The mainly photonic polariton (orange diamonds) remains delocalized, although presenting a slightly reduced PR.
Finally, in Fig.~\ref{fig:Finite size scaling}(c), a larger disorder strength $W/\omega_0=1$ is considered and in addition to the three eigenstates discussed above which are now semilocalized, we also show the scaling of the PR of a mainly photonic polariton with eigenfrequency $\omega_n^\mathrm{pol}/\omega_0=1.55$ (green circles) and of a fully dipolar dark state with eigenfrequency $\omega_n^\mathrm{pol}/\omega_0=0.7$ (black crosses).
In that context, one can see that all of the three previously discussed eigenstates show a PR that, after increasing at small sizes, converges to a constant value at larger sizes, confirming the fact that such semilocalized states are not extended.
Moreover, the mainly photonic polariton (green circles) still has a growing PR, showing that such a state having an eigenfrequency $\omega_n^\mathrm{pol}>\omega_0+W/2$ remains extended.
The eigenstates with frequencies $\omega_n^\mathrm{pol}<\omega_{q=0}^\mathrm{ph}$ newly allowed by the large value of the disorder strength are totally localized with a constant value of the PR around $1$, as can be seen from the black crosses in Fig.~\ref{fig:Finite size scaling}(c).
We note that we have also checked (not shown) that the photonic weight $\mathrm{Ph}(n)$ of the eigenstates is not affected when increasing the system size, and is independent of the number of dipoles $\mathcal{N}$.

\subsection{Level statistics analysis}

We propose here an analysis of the nature of the semilocalized states observed in our model by means of level spacing statistics. 
To this end, we follow Ref.~\cite{Oganesyan2007} and compute the distribution of level spacing ratios
\begin{equation}
    \tilde{r}_n = \mathrm{min}\left(r_n, \frac{1}{r_n}\right),
    \label{eq:level spacing ratio}
\end{equation}
with
\begin{equation}
    r_n=\frac{s_n}{s_{n-1}},
\end{equation}
where $s_n = \omega_{n+1}^\mathrm{pol} - \omega_n^\mathrm{pol} \geqslant 0$ is the level spacing, the eigenfrequencies $\omega_n^\mathrm{pol}$ being sorted in ascending order. 
The average ${\langle \tilde{r}_n \rangle}$ of such level spacing ratio over a given window of eigenstates $n$ has been shown to display a universal value according to the level statistics \cite{Atas13}.
Moreover, the average level spacing ratio is scale invariant at the critical point of an Anderson transition.
This allows one to monitor such transition between an extended phase, with statistics related to the random matrix ensembles, and a localized phase, with eigenfunctions exponentially localized on random sites, not overlapping so that the eigenvalues follow Poisson statistics \cite{Oganesyan2007,Tarquini2017,Torres-Herrera2019,Suntajs2021}.
In contrast to the standard level spacing distribution, the quantity \eqref{eq:level spacing ratio} has the advantage of not requiring any unfolding of the spectrum, since it is independent of the local density of states \cite{Oganesyan2007,Atas13}.

\begin{figure}
    \centering
    \includegraphics[width=\columnwidth]{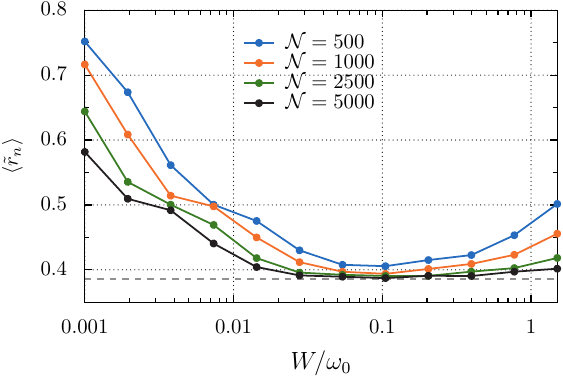}
    \caption{Average of the level spacing ratio ${\langle \tilde{r}_n \rangle}$ for eigenstates within the dark state band as a function of the disorder strength, for increasing sizes of the chain $\mathcal{N}$.
    The grey dashed line shows the exact value for Poisson statistics ${\langle \tilde{r}_n \rangle}^\mathrm{PE} = 2 \ln2 - 1 \simeq 0.39$.
    The cavity height $L_x/a=12$, and the data have been averaged over $100$ disorder realizations.}
    \label{fig:Level spacing}
\end{figure}

In Fig.~\ref{fig:Level spacing}, we show such one-parameter scaling for increasing number of dipoles $\mathcal{N}$, using the same parameters as in the previous subsection.
The averaging of the ratio \eqref{eq:level spacing ratio} has been performed with $1/10$ of the eigenstates of the spectrum centered around the middle of the dark state band in the case of an ordered chain, namely around the eigenfrequency $\omega_n^\mathrm{pol}/\omega_0 = 0.9981$.
The results were further averaged over $100$ disorder realizations.
As expected, there is no common intersection point between all the curves, demonstrating that there is no Anderson transition, as in usual 1d disordered systems.
Instead, the average level spacing ratio converges to the exact value computed for the localized phase with Poisson statistics, ${\langle \tilde{r}_n \rangle}^\mathrm{PE} = 2 \ln2 - 1 \simeq 0.39$ \cite{Atas13}, with diminishing disorder strength when the size of the system increases, as in the original 1d Anderson model \cite{Torres-Herrera2019}.
At small disorder strength, the average ratio goes to $1$, i.e., the value corresponding to an ordered system.
Therefore, as expected in $1$d \cite{Torres-Herrera2019}, there is no extended phase with statistics related to random matrix ensembles.

Interestingly, when increasing further the disorder strength, we observe in Fig.~\ref{fig:Level spacing} a rise of the average level spacing ratio, which corresponds to the semilocalization and hybridization of the dark states, as discussed in the main text.
The ratio ${\langle \tilde{r}_n \rangle}$ thus takes values between ${\langle \tilde{r}_n \rangle}^\mathrm{PE}$ and ${\langle \tilde{r}_n \rangle}=1/2$, which corresponds to semi-Poissonian statistics \cite{Bogomolny1999,Atas2013_2}.
Such behavior is in line with the results obtained in a disordered TC model where multifractal semilocalized states were found to follow statistics that
range from Poissonian to very close to semi-Poissonian, even at the thermodynamic limit \cite{Dubail22}.

Remarkably, however, in our multimode model we observe that increasing the size of the system leads the rise of the average level spacing ratio at large disorder to be flattened, so that the ratio converges to the Poisson value even at large disorder strength, as visible for the case of $\mathcal{N}=5000$ displayed by a black solid line in Fig.~\ref{fig:Level spacing}.
Hence, for larger systems, a larger disorder strength is required to push the semilocalized states away from a Poisson statistics.
This suggests that for a fixed, finite disorder strength $W/\omega_0$ with the parameters chosen here, namely $d/a=4$ and $L_x/a=12$, the semilocalized states in our model would just follow a Poisson statistics at the thermodynamic limit $\mathcal{N}\rightarrow\infty$, as usual localized states.

\subsection{Generalized participation ratio and multifractal analysis}

To ascertain whether or not the semilocalized states of our multimode model follow a multifractal structure as in single-mode models \cite{Dubail22}, here we conduct a multifractal analysis by analyzing the scaling with the system size $\mathcal{N}$ of the generalized participation ratio, which is defined as \cite{Mirlin_RevModPhys}
\begin{equation}
    \mathrm{PR}_{\mathfrak{q}}(n) = \frac{ \left(  \sum_{i=1}^{\mathcal{N}}|\Psi_i(n)|^2  \right)^{\mathfrak{q}}  }{ \sum_{i=1}^{\mathcal{N}}|\Psi_i(n)|^{2\mathfrak{q}}   } \underset{\mathcal{N}\to\infty}{\sim} \mathcal{N}^{\tau_{\mathfrak{q}}(n)}.
    \label{eq:Generalized PR real}
\end{equation}
While localized and delocalized eigenstates are characterized, respectively, by a multifractal exponent $\tau_{\mathfrak{q}}=0$ (with $\mathfrak{q}>0$) and $\tau_\mathfrak{q}=\mathfrak{q}-1$, any other behavior of $\tau_\mathfrak{q}$ as a function of $\mathfrak{q}$ implies multifractality \cite{Mirlin_RevModPhys}.
We extract the multifractal exponent $\tau_\mathfrak{q}$ from
a logarithmic linear regression of the generalized participation ratio \eqref{eq:Generalized PR real} as a function of the system size $\mathcal{N}$, that we average over a frequency window $\omega_n^\mathrm{pol}/\omega_0 \in [0.9\,;1.4]$ corresponding to hybridized, semilocalized states.
To minimize fluctuations, data obtained with system sizes $\mathcal{N}\in[10^2\,;10^3]$ and $\mathcal{N}\in~]10^3\,;3\times10^3]$ have been further averaged over, respectively, $100$ and $10$ disorder realizations.

\begin{figure}
    \centering
    \includegraphics[width=\columnwidth]{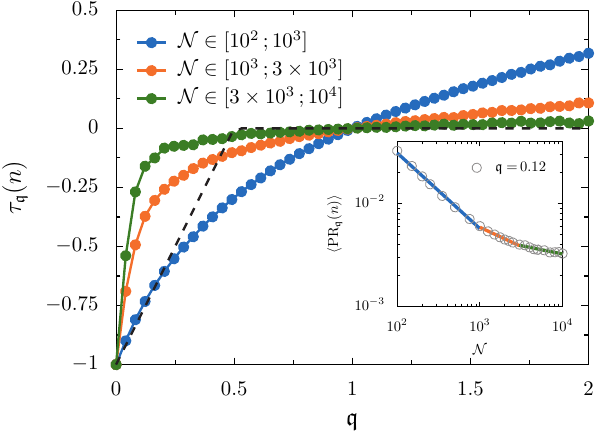}
    \caption{Multifractal exponent of the semilocalized states $\tau_\mathfrak{q}(n)$ [as defined in Eq.~\eqref{eq:Generalized PR real}] as a function of $\mathfrak{q}$.
    The exponents have been extracted according to the procedure explained in the text (see colored dots).
    The black dashed line represents the analytical result found for a simplified disordered Tavis-Cummings model \cite{Dubail22}.
    The inset exemplifies the scaling procedure for the value $\mathfrak{q}=0.12$, with the circles showing numerical data.
    The cavity height $L_x/a=12$ and the disorder strength $W/\omega_0=1$.}
    \label{fig:Multifractality}
\end{figure}

Figure \ref{fig:Multifractality} shows the resulting multifractal exponents extracted from different scaling procedures, considering only system sizes $\mathcal{N}\in[10^2\,;10^3]$ in blue, $\mathcal{N}\in[10^3\,;3\times 10^3]$ in orange, and $\mathcal{N}\in[3\times 10^3\,;10^4]$ in green.
In the inset we exemplify such scaling behavior for the value $\mathfrak{q}=0.12$, with numerical data represented as grey circles, and the linear regressions with solid lines.
We observe that when increasing the system size $\mathcal{N}$, the multifractal exponents $\tau_\mathfrak{q}$ are getting closer and closer to their value for localized states, namely $\tau_{\mathfrak{q}}=0$.
This drastically differs from the multifractal behavior found analytically at the thermodynamic limit in a simplified disordered TC model (see Eq.~(8) of Ref.~\cite{Dubail22}) which we show as a black dashed line in Fig.~\ref{fig:Multifractality}.
This suggests that the semilocalized states of our multimode model are not related to multifractal, nonergodic extended states at the thermodynamic limit.
We note that in a single-mode version of our model, namely considering the Hamiltonian \eqref{eq:Hamiltonian approximated} with only the photonic mode $n_z=1$, the exponent $\tau_\mathfrak{q}$ shows the same multifractal behavior as in the simplified disordered TC model considered in Ref.~\cite{Dubail22}.

To conclude, the analysis conducted in this Appendix suggests that the hybridized, semilocalized states present in our model through states localized in multiple, noncontiguous sites of the chain [see panels (c), (f), and (i) of Fig.~\ref{fig:Probability density for different disorder} where $\mathcal{N}=1000$], behave in the thermodynamic limit as usual localized states following Poissonian statistics.
This is in line with the results of our transport simulations discussed in the main text, where we show that disorder-enhanced transport, which is a consequence of the presence of hybridized, semilocalized states, vanishes for very large system sizes and is relevant at intermediate distances only (see Fig.~\ref{fig:Density plot transport versus site versus disorder}).
Therefore, in our multimode model, the semilocalization properties of the dark states that are hybridized through disorder fade out as the system size increases.



\section{Steady-state solution of the Lindblad master equation}
\label{sec:Stationary transport details}

In this Appendix, we detail our approach to find the steady-state solution of the Lindblad master equation \eqref{eq:Lindblad master equation} for the amplitudes of the dimensionless dipole moments $p_i = \langle b_i^{\phantom{\dagger}} + b_i^\dagger \rangle$ along the sites $i$ of the chain.
The latter quantity is directly proportional to the dipole moment at site $i$, $\mathfrak{p}_i = Q\sqrt{  {\hbar}/{2M\omega_0} }\,p_i$.
The associated power radiated in the far field by a dipole at site $i$ is then given by the classical Larmor formula \cite{Jackson2007}
\begin{equation}
    \mathcal{P}_i = \frac{2\omega_0^4}{3c^2}(\mathfrak{p}_i)^2 = \frac{4\pi a^2}{3}(k_0 a)^4I_0 \left( \frac{p_i}{\Omega_\mathrm{R}/\omega_0}\right)^2,
    \label{eq:Larmor}
\end{equation}
where $I_0=c{E_0}^2/8\pi$ is the driving field intensity.

In addition to $p_i$, we introduce a corresponding momentum $\pi_i = \mathrm{i}\langle b_i^{\phantom{\dagger}} - b_i^\dagger \rangle$, as well as the equivalent quantities for the photonic degrees of freedom, $p^\mathrm{ph}_{n_z} = \langle c_{n_z}^{\phantom{\dagger}} + c_{n_z}^\dagger \rangle$ and $\pi^\mathrm{ph}_{n_z} = \mathrm{i}\langle c_{n_z}^{\phantom{\dagger}} - c_{n_z}^\dagger \rangle$.
Using the identity $\langle \dot{O} \rangle = \mathrm{Tr}\left\{\dot{\rho}O\right\}$, valid for any operator $O$, one obtains from the master equation \eqref{eq:Lindblad master equation} the system of coupled first order ordinary differential equations
\begin{subequations}
    \begin{align}
        \dot{p}_i =&\; - \omega_i\pi_i - \frac{\gamma_i}{2} p_i +\sum_{n_z}\xi_{i n_z}p^{\mathrm{ph}}_{n_z} - \sum_{j\neq i} \Omega_{ij}\pi_j \\                                                           
        \dot{p}^{\mathrm{ph}}_{n_z} =&\; - \omega^{\mathrm{ph}}_{n_z}\pi^\mathrm{ph}_{n_z} - \frac{\kappa_{n_z}}{2} p^{\mathrm{ph}}_{n_z} - \sum_i \xi_{i n_z}p_i \\                      
        \dot{\pi}_i =&\; \omega_i p_i - \frac{\gamma_i}{2} \pi_i +  \sum_{n_z}\xi_{i n_z}\pi^{\mathrm{ph}}_{n_z} + \sum_{j\neq i} \Omega_{ij}p_j  \nonumber\\
        &\; + 2\Omega_\mathrm{R}f\left(t\right)\delta_{1,n} \\                
        \dot{\pi}^\mathrm{ph}_{n_z} =&\; \omega^\mathrm{ph}_{n_z}p^\mathrm{ph}_{n_z} - \frac{\kappa_{n_z}}{2} \pi^\mathrm{ph}_{n_z} - \sum_i \xi_{i n_z}\pi_i.
    \end{align}
    \label{eq:Lindblad system of equation}
\end{subequations}

Being interested in the stationary transport regime, we solve the above system for the steady-state solution using the complex representation $p_i = A_i \mathrm{e}^{\mathrm{i}\omega_\mathrm{d} t}$, and we recall that we consider a drive $f\left(t\right) = \mathrm{e}^{\mathrm{i}\omega_\mathrm{d} t}$.
The latter continuous periodic drive leads to a vanishing time-averaged dipole moment $\langle p_i \rangle_t$.
However, its time-averaged amplitude $\langle |p_i| \rangle_t$ is nonzero.
This procedure yields to steady-state solutions $\mathbf{P} = \left( |p_1|,|p_2|,\ldots,|p_\mathcal{N}|  \right)$ as
\begin{equation}
    \mathbf{P} = \mathcal{M}^{-1}\mathbf{D},
\end{equation}
where the $\mathcal{N}-$dimensional driving vector is given by $\mathbf{D} = \left( 2\Omega_\mathrm{R},0,\ldots,0 \right)$, and the $\mathcal{N}\times\mathcal{N}$ matrix $\mathcal{M}$ reads
\begin{align}
    \mathcal{M} =&  \left[ \left( \omega_\mathrm{d} - \mathrm{i} \frac{\gamma}{2} \right)\left( \omega_\mathrm{d} - \mathrm{i}\frac{\kappa}{2} \right)\mathbb{1}_\mathcal{N} - \mathcal{M}_\xi \left( \mathbb{1}_\mathcal{N} - \mathcal{D}_2   \right)  \mathcal{M}^{\mathrm{t}}_\xi \right] \nonumber \\           
   &\;\times \widetilde{\mathcal{M}}^{-1}_\mathrm{dp}  \left[  \frac{\omega_\mathrm{d} - \mathrm{i}\frac{\gamma}{2}}{\omega_\mathrm{d} - \mathrm{i}\frac{\kappa}{2}}\mathbb{1}_\mathcal{N} + \mathcal{M}_\xi \mathcal{D}_0  \mathcal{M}^{\mathrm{t}}_\xi  \right] - \widetilde{\mathcal{M}}_\mathrm{dp}.
\end{align}
In the above matrix equation, $ \mathbb{1}_\mathcal{N}$ is the $\mathcal{N}\times\mathcal{N}$ identity matrix, the matrix $\mathcal{M}_\xi$ is defined by its elements as $\left[\mathcal{M}_\xi\right]_{ij} = \xi_{ij}$, where $i,j = 1,\ldots,\mathcal{N}$, the diagonal matrices $\mathcal{D}_\beta = \mathrm{Diag}\left( \Omega^\beta_1, \Omega^\beta_2, \ldots, \Omega^\beta_\mathcal{N} \right)$, with the $\beta$-dependent function
\begin{equation}
    \Omega^\beta_{n_z} = \frac{ \left(\omega_{n_z}^{\mathrm{ph}}\right)^\beta } { (\omega_{n_z}^{\mathrm{ph}})^2 - \left( \omega_\mathrm{d} - \mathrm{i} \frac{\kappa}{2} \right)^2 },
\end{equation}
and finally the matrix $\widetilde{\mathcal{M}}_\mathrm{dp}$ is given by
\begin{equation}
    \widetilde{\mathcal{M}}_\mathrm{dp} = \mathcal{M}_\mathrm{dp} - \mathcal{M}_\xi \mathcal{D}_1 \mathcal{M}^\mathrm{t}_\xi,
\end{equation}
where $\mathcal{M}_\mathrm{dp}$ is defined by its elements as
\begin{equation}
    \left[\mathcal{M}_\mathrm{dp}\right]_{ij} = \omega_i \delta_{ij} + \Omega_{ij}\left( 1 - \delta_{ij} \right).
\end{equation}
We have checked (not shown) that such steady-state solutions $|p_i|$ are well recovered after a finite time when one directly solves the system \eqref{eq:Lindblad system of equation} using a numerical ODE solver.

\section{Effects of cavity losses}
\label{sec:Cavity losses}

\begin{figure}
    \centering
    \includegraphics[width=\columnwidth]{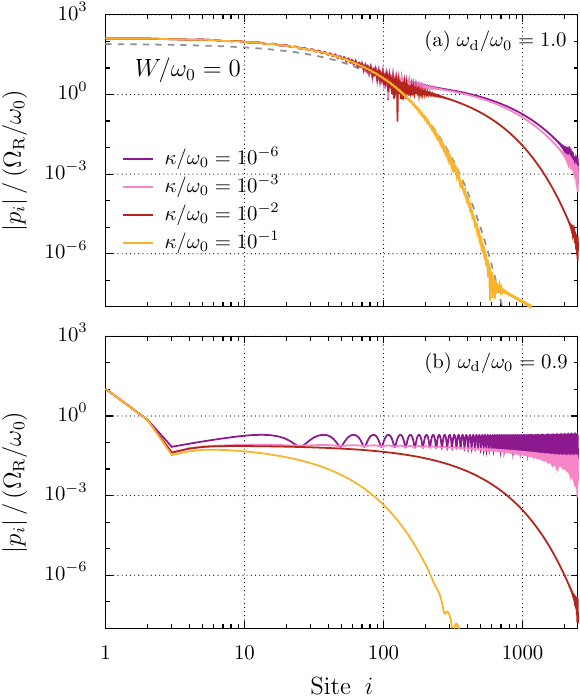}
    \caption{Steady-state amplitude of the dipole moment $|p_i|$ as a function of the dipole sites $i$ for different values of the cavity loss parameter $\kappa$, in the case of an ordered chain.
    Panels (a) and (b) correspond to the driving, respectively, of a dark state with $\omega_\mathrm{d}/\omega_0=1.0$ and of a mainly photonic polariton with $\omega_\mathrm{d}/\omega_0=0.9$.
    The grey dashed line in panel (a) represents the exponential decay of Eq.~\eqref{eq:Propagation analytical}.
    The dipole losses $\gamma/\omega_0=0.001$, and the cavity height $L_x/a=12$.}
    \label{fig:Cavity losses}
\end{figure}

In the main text, we fixed the cavity loss rate to $\kappa/\omega_0 = 10^{-3}$.
Here, we motivate this choice by comparing in Fig.~\ref{fig:Cavity losses} the propagation characteristics of an ordered chain for different values of cavity loss, from $\kappa/\omega_0 = 10^{-6}$ to $\kappa/\omega_0 = 10^{-1}$.
The amplitude of the steady-state dipole moment $|p_i|$ in units of the reduced Rabi frequency $\Omega_\mathrm{R}/\omega_0$ is shown as a function of the dipole sites $i$ along the chain.
Two driving frequencies $\omega_\mathrm{d}/\omega_0 = 1.0$ and $\omega_\mathrm{d}/\omega_0 = 0.9$ are shown in Figs.~\ref{fig:Cavity losses}(a) and \ref{fig:Cavity losses}(b), respectively.
The dipole losses are fixed to $\gamma/\omega_0=0.001$, while the cavity height is chosen as $L_x/a=12$.

In both panels, the qualitative behavior is substantially the same for cavity losses $\kappa/\omega_0=10^{-6}$ (purple lines) and $\kappa/\omega_0=10^{-3}$ (pink lines), justifying the choice made in the main text.
We observe that considering larger cavity losses $\kappa/\omega_0=10^{-2}$ (red lines) does not qualitatively change the results when driving a dark state in panel (a), showing the robustness of dark state propagation against cavity losses.
As expected, however, the propagation of the mainly photonic polariton shown in Fig.~\ref{fig:Cavity losses}(b) is more affected by the increase of cavity losses.
Crucially, by increasing further the cavity loss rate to $\kappa/\omega_0 = 10^{-1}$ (yellow lines), the second exponential regime, that is, the cavity-enhanced transport, completely disappears in Fig.~\ref{fig:Cavity losses}(a), showing that the system is not anymore in the strong-coupling regime.
The dark state propagation is then very well described by the exponential regime mediated solely by the nearest-neighbor quasistatic dipole-dipole coupling, given by the analytical result of Eq.~\eqref{eq:Propagation analytical} and shown as a dashed grey line.
In the case of the mainly photonic polariton in Fig.~\ref{fig:Cavity losses}(b), we observe similar results, namely that the long-range transport starts to become suppressed for cavity losses $\kappa/\omega_0=10^{-1}$.
This transition from the strong- to the weak-coupling regime is expected, since the latter large cavity loss rate is of the order of the Rabi splitting frequency, $\Omega_\mathrm{RS} \simeq 0.12$.
We note that similar conclusions can be drawn when the dipole losses are fixed to a larger value.

Finally, we recall that the cavity quality factor $\mathcal{Q}$ is typically of the order of the inverse of the cavity loss rate in units of the average dipole frequency $\omega_0/\kappa$.
This implies that the effects discussed in this work should be observed with cavity quality factors as low as $\mathcal{Q} \sim 10^2$.


\section{Absorption spectra from transport computations}
\label{sec:Absorption}

\begin{figure}[tb]
    \centering
    \includegraphics[width=\columnwidth]{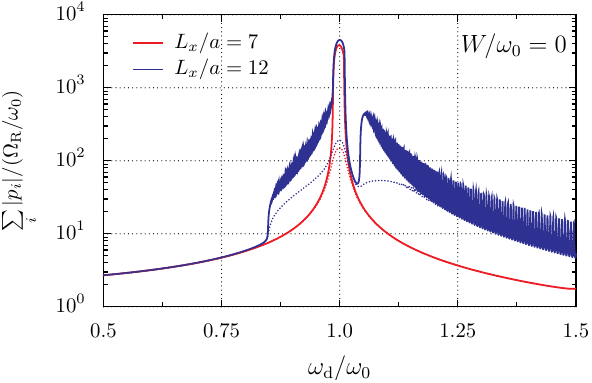}
    \caption{Absorption spectrum as the sum of $\mathcal{N}=1000$ dipole moment amplitudes $|p_i|$ in units of the reduced Rabi frequency $\Omega_\mathrm{R}/\omega_0$, as a function of the reduced driving frequency $\omega_\mathrm{d}/\omega_0$.
    Results are shown without disorder for two different cavity heights $L_x/a = 7$ (red lines, weak coupling) and $L_x/a = 12$ (blue lines, strong coupling), and two different dipole losses $\gamma/\omega_0 = 0.001$ (solid lines) and $\gamma/\omega_0 = 0.02$ (dotted lines).}
    \label{fig:Absorption ordered different Lx}
\end{figure}

In this Appendix, we complement the discussion on transport of Sec.~\ref{sec:Transport} by presenting absorption spectra obtained from our driven-dissipative computations.
To this end, we consider that the sum of all the steady-state dipole moment amplitudes, $\sum_{i=1}^\mathcal{N} |p_i|$, is at first order proportional to the electric field absorbed by the dipoles \cite{Sturges2020}.

\subsection{Ordered chain}

We begin with the case of an ordered chain, and show in Fig.~\ref{fig:Absorption ordered different Lx} such absorption spectrum for cavity heights $L_x/a = 7$ (blue lines) and $L_x/a = 12$ (red lines). 
Two different dipole losses $\gamma/\omega_0 = 0.001$ (solid lines) and $\gamma/\omega_0 = 0.02$ (dotted lines) are considered.

While in the weak coupling regime, $L_x/a=7$, the absorption spectrum is bell-shaped and centered around the dark state eigenfrequencies at $\omega_\mathrm{d}/\omega_0=1.0$, in the strong-coupling regime, $L_x/a=12$, a rich structure appears.
In addition to the central peak, which is slightly higher, we observe high absorption for driving frequencies associated to hybridized polaritons (see the dispersion in Fig.~\ref{fig:Dispersion Fourier}), namely the bottom of the LP branch, starting around $\omega_\mathrm{d}/\omega_0=0.85$, and the UP branch, visible as a second peak starting around $\omega_\mathrm{d}/\omega_0=1.05$ and separated from the LP branch by a gap.
The fast oscillations are peaks corresponding to the different eigenfrequencies of the system, visible one by one due to the steep slope in the dispersion in these frequency regions.
We can hence precisely recover the eigenmodes from this driven-dissipative scenario.
However, the absorption spectrum of Fig.~\ref{fig:Absorption ordered different Lx} does not reflect the transport properties along the whole chain, the sum over the amplitudes $|p_i|$ being obviously dominated by the first few terms.
It then provides information mostly about the short-range propagation.
This explains why the absorption is maximal for the dark eigenstates, and decreases when the driving frequency rises to more photonic eigenstates.
Indeed, as discussed in Sec.~\ref{sec:Transport}, the dark states present the most efficient transport at short distances.

By increasing the dipole losses, from the solid to the dotted lines, the central peak is drastically reduced and the full width at half maximum is increased, as expected.
The polaritons at the bottom of the LP and UP branches are less distinct from the dark states of the central peak, the latter being the most affected by dipole losses.
Crucially, however, the two bands are still distinguishable and separated by a gap despite the larger dipole losses, indicating that the system is still in the strong-coupling regime.

\subsection{Disordered chain}

\begin{figure}[tb]
    \centering
    \includegraphics[width=\columnwidth]{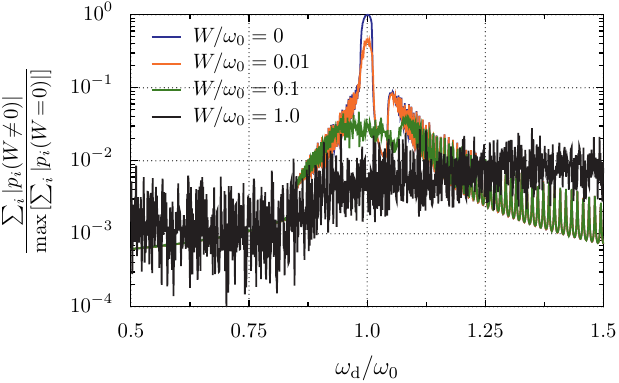}
    \caption{Absorption spectrum for different disorder strengths $W$, normalized by the maximum of the one obtained without disorder.
    The results are shown for a chain of $\mathcal{N}=500$ dipoles, with a cavity height $L_x/a=12$, dipole losses $\gamma/\omega_0=0.001$, and have been averaged over $100$ disorder realizations.}
    \label{fig:Absorption different disorder}
\end{figure}

The case of a disordered chain is shown in Fig.~\ref{fig:Absorption different disorder}, where the sum of the amplitudes of the dipole moments along the chain for a given disorder strength, $\sum_{i=1}^\mathcal{N}|p_i(W\!\neq\!0)|$, has been normalized by the maximum of the one found without disorder, $\mathrm{max}[\sum_{i=1}^\mathcal{N}|p_i(W\!=\!0)|]$.
We choose a cavity height $L_x/a=12$ and we consider small dipole losses $\gamma/\omega_0=0.001$.
By increasing the disorder strength from $W/\omega_0=0$ (blue line) to $W/\omega_0=0.01$ (orange line), and to $W/\omega_0=0.1$ (green line), the absorption is more and more suppressed for the driving frequencies corresponding to dark states, in a growing frequency window around $\omega_\mathrm{d}/\omega_0=1.0$.
This is in agreement with what we observe in Sec.~\ref{sec:Spectrum}, namely that the effective dark state band grows with the disorder strength $W$ around $\omega_0$, while the polaritons outside the dark state band, with $\omega_n^{\mathrm{pol}} \gtrsim \omega_0 + W/2$, are not affected by disorder.
The polaritons hence profit from this cavity-protection effect due to their eigenfrequencies being far from $\omega_0$.

With a disorder strength $W/\omega_0=1$ (black line in Fig.~\ref{fig:Absorption different disorder}), the dark state band includes now frequencies up to $1.5\,\omega_0$.
Due to the disorder-induced hybridization mechanism, eigenstates at frequencies around $1.5\,\omega_0$ have gone from almost purely photonic polaritons to semilocalized polaritonic states with a significant dipolar part.
This leads their absorption to become larger than the one without disorder, the dipolar part allowing them to be more easily excited and to acquire a better short-range propagation mediated by the quasistatic dipole-dipole coupling.
Importantly, we thus observe here a disorder-enhanced short-range transport, from almost fully photonic modes acquiring a dipolar part from the increased disorder strength, and inheriting efficient short-range dipolar transport.


\bibliography{refs}
\end{document}